\documentclass{iopjournal}

\usepackage{color}
\usepackage{amsmath, amssymb,amsfonts,bbold}
\usepackage{soul}
\usepackage[normalem]{ulem}
\usepackage{cancel}

\DeclareMathOperator{\Tr}{Tr}

\usepackage{comment}
\usepackage{graphicx}
\usepackage{subcaption}
\usepackage{tikz}
\usetikzlibrary{decorations.pathmorphing}
\usepackage{braket}

\usepackage{ragged2e}

\usepackage[
  style=phys,
  backend=biber,
  biblabel=brackets,
  articletitle=false,
]{biblatex}

\allowdisplaybreaks

\begin{document}
\justifying
\pagestyle{empty}

\title{Steady-State Noise Signatures of Lindbladian Exceptional Points}
\author{Shihang Pan$^1$, Gianmichele Blasi$^{1,2}$ and G\'eraldine Haack$^1$}

\affil{$^1$Department of Applied Physics, University of Geneva, 1211 Geneva, Switzerland}\\
\affil{$^2$Instituto de Física Interdisciplinar y Sistemas Complejos IFISC (CSIC-UIB), E-07122 Palma de Mallorca, Spain}

\begin{abstract}
\justifying
 Exceptional points (EPs) are non-Hermitian degeneracies at which two or more eigenvalues and their corresponding eigenvectors coalesce. In open quantum systems, exceptional points can arise in the Lindbladian governing the dissipative dynamics. Their signatures have so far been mainly identified in finite-time observables, such as transient currents, while steady-state average currents generally provide no direct evidence of the underlying exceptional-point structure. In this work, we demonstrate that signatures of Lindbladian EPs can nevertheless be accessed in the steady-state regime through current noise. 
 We derive general expressions for current correlation functions within a Lindblad master-equation framework and show, in particular, how exceptional points affect their behaviour as a function of the time delay.
 We illustrate these results with the paradigmatic example of two interacting qubits coupled to two reservoirs, where the steady-state noise clearly distinguishes overdamped, underdamped, and critical regimes. Our results establish current correlation functions as a steady-state probe of Lindbladian EPs in open quantum systems.
\end{abstract}

\section{Introduction}

In recent years, the study of non-Hermitian physics has emerged as a powerful framework for understanding open quantum systems \cite{el2018non, bender2007making, rotter2009non, moiseyev2011non} — systems that interact with their environment and hence cannot be described by traditional Hermitian quantum mechanics \cite{breuer2007theory, rivas2012open, schaller2014open}. A particularly striking feature of non-Hermitian operators is the occurrence of exceptional points (EPs) \cite{heiss2012physics, minganti2019quantum, miri2019exceptional, hatano2019exceptional}, degeneracies at which both the eigenvalues and eigenvectors of a system coalesce \cite{kato2013perturbation}. Mathematically, they mark the breakdown of diagonalizability and are associated with non-trivial Jordan block structures \cite{horn1985matrix} in the spectral decomposition of non-Hermitian matrices, Hamiltonians or Lindbladians \cite{ashida2020non, wang2021topological, Hu2025}.

These singularities open the way to possibly rich phenomena. While EPs of non-Hermitian Hamiltonians give rise to specific energy spectrum features (which can be associated in certain cases to phase transitions and topological properties) \cite{heiss1998collectivity,muller2008exceptional, shen2018topological, sun2024encircling, chen2022decoherence, abbasi2022topological}, EPs of Lindbladians influence the dynamics of open quantum systems (OQS) by modifying the decay quantum channels and their associated decay rates towards the steady state. When investigating average observables in OQS, like out-of-equilibrium currents (particle, charge and heat currents), signatures of EPs have been evidenced at finite times, e.g. in the transient regime. Their dynamical impact is profound: time evolutions acquire polynomial corrections to exponential decays, leading to critical behaviors \cite{minganti2019quantum, minganti2020hybrid, am2015exceptional, khandelwal2021signatures, zhou2023accelerating, zelle2024universal, khandelwal2025emergent, Solow2025}. Recent progress confirms that signatures of LEPs (Jordan blocks leading to polynomial time
prefactor) can be directly exploited to control open-system dynamics, providing a route to dissipative
entanglement generation~\cite{khandelwal2024chiral, li2023speeding} or linking LEPs to quantum information tasks~\cite{kumar2022optimized, arkhipov2023dynamically}.
They have also been related to parameter estimation~\cite{am2015exceptional, wiersig2020robustness, chen2019sensitivity, wiersig2014enhancing, zhang2019quantum} and thermodynamic performance for quantum thermal machines~\cite{zhang2022dynamical, bu2023enhancement} and quantum metrology applications \cite{Konye2024}.
This broad relevance is further supported by recent results showing that Liouvillian exceptional points can survive beyond weak-coupling Markovian descriptions, emerging instead as robust features of open quantum dynamics~\cite{khandelwal2025emergent,lin2025non}. 


Despite all these studies, finding relevant signatures of EPs in observables accessible in realistic experiments, \textit{i.e.} able to access sufficiently short time dynamics to evidence the time-polynomial dependence associated with EPs, is a current open and challenging question. In this work, we tackle this problem by investigating the role of Lindbladians EPs (LEPS) in correlation functions of transport observables within a master equation approach. Remarkably, we find that LEPs exhibit signatures in the steady-state regime in the current-noise function, opening new experimental possibilities to witness the presence and effects of these genuine non-Hermitian properties. We illustrate these findings with a specific example of two interacting qubits embedded within a two-terminal transport setup. All our results are analytical, allowing for a fundamental understanding of the role played by Jordan blocks in correlation functions.

The manuscript is organized as follows. In Sec.~\ref{sec:framework}, we consider a generic model for an multi-qubit OQS within a master equation approach and introduce the corresponding Lindbladian. In Sec.~\ref{sec:signatures}, we first recall the definitions of transport observables of interest to this work, average current and two-point current correlation function. This allows us to provide analytical signatures of Jordan blocks and LEPs in the steady-state regime. In Sec.~\ref{sec:two-qubit}, we illustrate these general results for the specific case of two interacting qubits embedded in a two-terminal device. Both environments are biased in temperature and/or voltage, giving rise to out-of-equilibrium currents flowing through the two qubits. We demonstrate the existence of analog underdamped, critical and overdamped regime in the noise as a function of the time delay. We conclude this work with a general discussion and perspectives in Sec.~\ref{sec:ccl}.

\section{Master equation framework for the dynamics of OQS} \label{sec:framework} 
In this section, we introduce the general framework of master equations for assessing the quantum dynamics of a quantum system in interaction with one or several Markovian environments. We present the most general decomposition of the Lindbladian.

\subsection{Master equation approach}
We assume a generic system consisting of $N$ qubits labelled $j = 1, \ldots, N$, individually tunnel-coupled to Markovian reservoirs at equilibrium, $\alpha = 1,\ldots, N$. These qubits may be interacting or not, we discuss specific cases below. The total Hamiltonian takes the form:
\begin{equation}
	H = H_S + H_B + H_{SB} \label{eq:ham_tot}
\end{equation}
with $H_S$ the system Hamiltonian, $H_B$ the Hamiltonians for the reservoirs and $H_{SB}$ the system-bath tunneling Hamiltonians,
\begin{align}
	H_{B} &= \sum_{\alpha = 1}^{N} \sum_{k}  \omega_{k \alpha} c^{\dagger}_{k \alpha} c_{k \alpha}\, \label{eq:HB} \\
	H_{SB} &= \sum_{j = 1}^{N} \sum_{\alpha = 1}^{N} \sum_{k} \left( \gamma_{j k \alpha} \sigma_{+}^{\left( j \right)} c_{k \alpha} + \gamma_{j k \alpha}^{*} \sigma_{-}^{\left( j \right)} c_{k \alpha}^{\dagger} \right). \label{eq:HSB}
\end{align}
The bath Hamiltonian corresponds to a sum of  bosonic or fermionic baths at equilibrium labelled $\alpha$, with $\omega_{k \alpha}$ the energy of mode $k$ in reservoir $\alpha$, and $c^{\dagger}_{k \alpha}, c_{k \alpha}$ the corresponding creation and annihilation operators that satisfy commutation $\left[ c_{k \alpha}, c^{\dagger}_{k' \beta} \right] = \delta_{\alpha \beta} \delta_{k k'}$ (bosonic) or anti-commutation $\left\{ c_{k \alpha}, c^{\dagger}_{k' \beta} \right\} = \delta_{\alpha \beta} \delta_{k k'}$ (fermionic) relations. For the system Hamiltonian $H_S$, we assume a generic one with flip-flop type interaction Hamiltonian among nearest-neighbor qubits:
\begin{equation}
    H_{S} = \sum_{j = 1}^{N} \epsilon_{j} \sigma_{+}^{\left( j \right)} \sigma_{-}^{\left( j \right)} + \sum_{\langle i, j \rangle} g_{i j} \left( \sigma_{+}^{\left( i \right)} \sigma_{-}^{\left( j \right)} + \sigma_{+}^{\left( j \right)} \sigma_{-}^{\left( i \right)} \right). \label{eq:ham_syst}
\end{equation}
The bare energy $\epsilon_{j}$ of qubit $j$ is assumed to be the largest energy scale of the problem. In general, the Bohr frequencies $\epsilon_{i} - \epsilon_{j}$ (also commonly known as the energy detuning), as well as the interaction strength $g_{i j}$ can be of the order or smaller than the coupling strength $\gamma_{j k \alpha}$. This assessment is crucial to determine the valid form of a Lindblad equation, it will determine whether \textit{local} or \textit{global} jump operators enter the dynamics, corresponding to jump operators acting onto the canonical or energy eigenbasis respectively \cite{brask2015autonomous, hofer2017markovian, khandelwal2020critical}. We emphasize that the general framework and analysis we present below are valid in both cases. Hence, for the sake of clarity and accessibility, we assume in the rest of the manuscript that the dynamics induced by the presence of the baths is correctly captured by \textit{local} jump operators, the operators $\sigma_{\pm}^{\left( j \right)}$ in the case of a multi-qubit system considered here. This is correct if all Bohr frequencies and coupling strengths $g_{j k}$ are smaller than the coupling strength between system and baths. Under this assumption, the Lindblad equation for the density operator of the system $\rho$ takes the form:
\begin{align}
    \dot{\rho} \left( t \right)
	&= -i \left[ H_{S}, \rho \left( t \right) \right] + \sum_{\alpha} \sum_{j = 1}^{N} \left( \gamma_{j \alpha}^{+} \mathcal{D} \left[ \sigma_{+}^{ \left( j \right)} \right] + \gamma_{j \alpha}^{-} \mathcal{D} \left[ \sigma_{-}^{\left( j \right)} \right] \right) \rho \left( t \right) \label{eq:ME} \\
    &\equiv \mathcal{L} \rho \left( t \right). \label{eq:Lindbladian}
\end{align}
Here, the superoperators $\mathcal{D} \left[ X \right] \cdot = X \cdot X^{\dagger} - \frac{1}{2} \left\{ X^{\dagger} X, \cdot \right\}$ correspond to the dissipators with the definition $\left\{ A, B \right\} = A B + B A$ and $\mathcal{L}$ is the Lindbladian (equivalently Lindbladian in our context), of interest for the rest of this work. The general solution of this first-order differential equation is simply:
\begin{equation}
    \rho \left( t \right) = e^{\mathcal{L} t} \rho_{0}, \label{eq:sol_rho}
\end{equation}
with $\rho_{0} \equiv \rho \left( t = 0 \right)$ the density operator at initial time. Eq.~\eqref{eq:sol_rho} indicates that the dynamics of $\rho$ is fully controlled by the superoperator $\mathcal{L}$. In the next subsection, we discuss the spectral properties of the Lindbladian $\mathcal{L}$ to be exploited in the rest of the manuscript.


\subsection{Spectral decomposition of the Lindbladian}

We investigate the spectral properties of the Lindbladian by employing standard vectorization techniques. For an $N$-dimensional system, the density matrix $\rho$ becomes a vector of dimension $N^{2}$, while the Lindbladian takes the form of a matrix $L$ of dimension $N^{2} \times N^{2}$ with $N^2$ eigenvalues. We note the corresponding spectrum $\sigma \left( L \right) = \left\{ \lambda_{1}, \ldots, \lambda_{N^2} \right\}$ and recall that a non-Hermitian matrix can be diagonalizable if and only if a set of linearly independent eigenvectors can be found. Due to dissipation induced by the jump terms, the matrix $L$ is non-Hermitian with complex eigenvalues in general, $\sigma \left( L \right) \in \mathbb{C}$. We recall that hermiticity of the density operator implies the coefficients of the characteristic polynomial of $\mathcal{L}$ to be real-valued. Hence, the eigenvalues $\lambda_{i}$ are either real or come in complex conjugate. The proof is shown in App.~\ref{sec:appendix}.

For the most general situation, we consider a generic type of spectrum $\sigma \left( L \right)$, made of complex eigenvalues, real eigenvalues and degenerate eigenvalues corresponding to EPs. More specifically, we assume $2 p$ complex eigenvalues, $\theta_{1}, \theta_{1}^{*}, \cdots, \theta_{p}, \theta_{p}^{*}$, $m$ identical eigenvalues $\gamma$ corresponding to an $m$-th order exceptional point, and $N^{2} - 2 p - m$ real eigenvalues denoted $\lambda_{1}, \cdots, \lambda_{N^{2} - 2 p - m}$. Without loss of generality, we assume a one-dimensional kernel for $L$, corresponding to the eigenvalue $\lambda_{1} = 0$ and a unique steady state \cite{nigro2019uniqueness}. We then suppose $\sigma_{1}, \cdots, \sigma_{N^{2} - 2p - m}$ and $\rho_{1}, \cdots, \rho_{N^{2} - 2p - m}$ are left- and right- eigenmatrices corresponding to the real eigenvalues; $\mu_{1}, \mu_{1}^{\dagger}, \cdots, \mu_{p}, \mu_{p}^{\dagger}$ and $\xi_{1}, \xi_{1}^{\dagger} \cdots, \xi_{p}, \xi_{p}^{\dagger}$ are left- and right- eigenmatrices corresponding to the complex eigenvalues $\theta_{1}, \theta_{1}^{*}, \cdots, \theta_{p}, \theta_{p}^{*}$; $\zeta_{1}, \cdots, \zeta_{m}$ and $\eta_{1}, \cdots, \eta_{m}$ are generalized left- and right-eigenmatrices corresponding to the $m$ identical eigenvalues $\gamma$. All the left- and right- eigenmatrices are orthonormal to each other. The decomposition of $\mathcal{L}$ takes the form
\begin{equation}
	\mathcal{L} = \sum_{i = 1}^{N^{2} - 2p - m} \lambda_{i} \rho_{i} \sigma_{i}^{\dagger} + \sum_{i = 1}^{p} \left( \theta_{i} \xi_{i} \mu_{i}^{\dagger} + \theta_{i}^{*} \xi_{i}^{\dagger} \mu_{i} \right) + \gamma \sum_{i = 1}^{m} \eta_{i} \zeta_{i}^{\dagger} + \sum_{i = 1}^{m - 1} \eta_{i} \zeta_{i + 1}^{\dagger}\,. \label{eq:spectrum}
\end{equation}

Written in its eigenbasis, the Lindbladian takes the Jordan normal form
\begin{equation}
    L =
    \begin{pmatrix}
        \Lambda_{N^{2} - m} & \\
        & J_{m} \left( \gamma \right)
    \end{pmatrix}
\end{equation}
where $\Lambda_{N^{2} - m}$ is a diagonal matrix of order $N^{2} - m$ containing all eigenvalues except $\gamma$,
\begin{equation}
    \Lambda_{N^{2} - m} = \operatorname{diag} \left( \lambda_1 = 0, \lambda_2, \ldots, \lambda_{N^2-2p-m}, \theta_1, \theta_1^*, \ldots, \theta_p, \theta_p^* \right)
\end{equation}
and $J_m(\gamma)$ is the Jordan block of order $m$ associated with the
exceptional eigenvalue $\gamma$, defined as
\begin{equation}
J_m(\gamma) = \gamma  \mathbb{1}_m + N_m,
\end{equation}
with $\mathbb{1}_m$ the $m \times m$ identity matrix and $N_m$ a nilpotent matrix
with ones on the first superdiagonal and zeros elsewhere, i.e. $(N_m)_{ij} = \delta_{i,j-1}$.\\

As mentioned previously in Eq.~\eqref{eq:sol_rho}, these spectral properties can be used to express the solution $\rho \left( t \right)$, as a function of the eigenvalues and eigenmatrices of $\mathcal{L}$
\begin{equation}
    \rho \left( t \right) = \rho_{ss} + \sum_{i = 2}^{N^{2} - 2p - m} b_{i} e^{\lambda_{i} t} \rho_{i} + \sum_{i = 1}^{p} \left( c_{i} e^{\theta_{i} t} \xi_{i} + c_{i}^{*} e^{\theta_{i}^{*} t} \xi_{i}^{\dagger} \right) + e^{\gamma t} \sum_{i = 1}^{m} d_{i} \left( \sum_{j = 1}^{i} \frac{t^{i - j}}{\left( i - j \right) !} \eta_{j} \right)
\end{equation}
with $b_{i} = \langle \sigma_{i}^{\dagger} \rho_{0} \rangle$, $c_{i} = \langle \mu_{i}^{\dagger} \rho_{0} \rangle$, and $d_{i} = \langle \zeta_{i}^{\dagger} \rho_{0} \rangle$ the Hilbert-Schmidt product of the initial density operator with the $i$-th left eigenmatrix. By normalization, we choose $\sigma_{1} = \mathbb{I}$ such that $\rho_{ss} = \rho_{1}$. This equation explicitly shows that steady-state density operators $\rho_{ss}$ and observables that depend only on $\rho_{ss}$ cannot exhibit genuine non-Hermitian properties of the evolution matrix, the Lindbladian. For those quantities, signatures of EPs have to be investigated at finite times, \textit{i.e.} in the transient regime, see Refs.~\cite{khandelwal2021signatures, bourgeois2026transport}. In contrast, we show below that higher-order correlation functions provide unexplored ways to witness non-Hermitian properties in the steady-state regime. To this end, we first recall how to define transport quantities within a master equation approach.

\section{Transport observables} \label{sec:signatures}

In this section, we provide general formula for calculating the average particle current and the corresponding two-point current correlation function within a master equation approach. All expressions are expressed in terms of the Lindbladian and its spectral properties. 

\subsection{Full counting statistics}
Within a master equation approach, averaged transport observables can be calculated. Explicit and general formula were derived in recent works, see Refs.~\cite{blasi2024exact,landi2024current} and references therein for older works discussing the average current. Assuming the validity of the master equation Eq.~\eqref{eq:ME}, one can define the current superoperator in reservoir $\alpha$:
\begin{equation}
    \mathcal{I}_{\alpha} \bullet = \sum_{j} \left( \gamma_{j\alpha}^{+}\sigma_+^{(j)} \bullet \sigma_-^{(j)} - \gamma_{j\alpha}^{-}\sigma_-^{(j)} \bullet \sigma_+^{(j)} \right) \label{eq:current_supop}
\end{equation}
from which the averaged particle current into reservoir $\alpha$ can be calculated:
\begin{equation}
   I_\alpha (t) = \Tr\{\mathcal{I}_\alpha \rho(t)\} = \Tr\{ \mathcal{I}_\alpha e^{\mathcal{L} t} \rho_0 \}.
\end{equation}
At steady state corresponding to the infinite time limit, only $\lambda_1 =0$ contributes to the solution. We therefore obtain: 
\begin{equation}
    I_{\alpha}^{ss} = \lim_{t \to \infty} I_{\alpha} \left( t \right) = \Tr \left\{ \mathcal{I}_{\alpha} \rho_{ss} \right\}.
\end{equation}
with $\rho_{ss}$ representing the steady-state value of the currents into reservoir $\alpha$. Two-point current correlation functions can also be defined within a master equation. We first recall their definition:
\begin{equation}
    S_{\alpha \alpha'} \left( t, t' \right) = \langle I_{\alpha} \left( t \right) I_{\alpha'} \left( t' \right) \rangle - \langle I_{\alpha} \left( t \right) \rangle \langle I_{\alpha'} \left( t' \right) \rangle,
\end{equation}
where $\langle \cdot \rangle$ denotes the quantum statistical average over the equilibrium distribution function of the reservoir. Within a master equation approach, it was shown in Ref.~\cite{blasi2024exact} that the noise is given by the following expression:
\begin{align}
    S_{\alpha\alpha'}(t, t')
    &= \delta_{\alpha\alpha'}\delta(t'-t)\Tr(\mathcal{A}_{\alpha}e^{\mathcal{L}t}\rho) - \Tr(\mathcal{I}_{\alpha}e^{\mathcal{L}t}\rho_{0})\Tr(\mathcal{I}_{\alpha'}e^{\mathcal{L}t'}\rho) \notag \\
    & \quad + \theta(t'-t)\Tr(\mathcal{I}_{\alpha'}e^{\mathcal{L}(t'-t)}\mathcal{I}_{\alpha}e^{\mathcal{L}t}\rho) + \theta(t-t')\Tr(\mathcal{I}_{\alpha}e^{\mathcal{L}(t-t')}\mathcal{I}_{\alpha'}e^{\mathcal{L}t'}\rho), \label{eq:noise}
\end{align}
with 
\begin{equation}
    \mathcal{A}_\alpha \bullet = \sum_{j} \left( \Gamma_{j\alpha}^{+}\sigma_+^{(j)} \bullet \sigma_-^{(j)} + \Gamma_{j\alpha}^{-}\sigma_-^{(j)} \bullet \sigma_+^{(j)} \right) \label{eq:activity}
\end{equation}
the so-called activity that corresponds to the rate of quantum jumps between system and reservoir. Note the $+$ sign to be contrasted with the $-$ sign in the expression of the current superoperator, Eq.~\eqref{eq:current_supop}. Equation~\eqref{eq:noise} clearly shows that the noise at finite times will feature genuine non-Hermitian properties through the role of $\mathcal{L}$. The question we address is whether signatures will be present in the steady-state regime. To this end, we derive the long-time limit expression of Eq.~\eqref{eq:noise}, also available in Ref.~\cite{blasi2024exact}. At long times, the two-point current correlation function depends only on a single time, the time delay $\tau = t'-t$. It reads: 
\begin{align}
S_{\alpha\alpha'}^{(ss)}(\tau) 
    &= \lim_{t\to\infty}S_{\alpha\alpha'}(t, t+\tau) \notag \\
    &= \delta_{\alpha\alpha'}\delta(\tau)\Tr(\mathcal{A}_{\alpha}\rho_{ss})-\Tr(\mathcal{I}_{\alpha}\rho_{ss})\Tr(\mathcal{I}_{\alpha'}\rho_{ss}) \notag \\
    & \quad + \theta(\tau)\Tr(\mathcal{I}_{\alpha'}e^{\mathcal{L}\tau}\mathcal{I}_{\alpha}\rho_{ss})+\theta(-\tau)\Tr(\mathcal{I}_{\alpha}e^{-\mathcal{L}\tau}\mathcal{I}_{\alpha'}\rho_{ss}).
\end{align}

While the steady-state activity and average currents will not depend on the spectral properties of $\mathcal{L}$, the last two terms will. In the next section, we investigate in detail the behaviour of the steady-state noise at EP and away from EP. This allows us to derive non-ambiguous signatures of LEPs in the steady-state regime.

\subsection{Explicit expressions of transport observables}
We then investigate in detail the behavior of the currents in transient regime and the steady-state noise at and away from exceptional points. Exploiting the spectral decomposition of the Lindbladian $\mathcal{L}$ in Eq.~\eqref{eq:spectrum}, we get
\begin{align}
    I_{\alpha} \left( t \right)
    &= I_{\alpha}^{ss} + \sum_{i = 2}^{2^{2 N} - 2 p - m} b_{i}^{\alpha} e^{\lambda_{i} t} + 2 \sum_{i = 1}^{p} e^{\Re \left( \theta_{i} \right) t} \left( \Re \left( c_{i}^{\alpha} \right) \cos \left[ \Im \left( \theta_{i} \right) t \right] - \Im \left( c_{i}^{\alpha} \right) \sin \left[ \Im \left( \theta_{i} \right) t \right] \right) \notag \\
    & \quad + e^{\gamma t} \sum_{i = 1}^{m} \sum_{j = 1}^{i} \frac{d_{i j}^{\alpha}}{\left( i - j \right) !} \tau^{i - j}. \label{eq:currents}
\end{align}
All coefficients in Eq.~\eqref{eq:currents} calculated from the overlap of the density operator at initial time with the left-eigenmatrices, which are dependent on set of parameters, are defined as the following
\begin{align}
	b_{i}^{\alpha} &= \Tr \left\{ \sigma_{i}^{\dagger} \rho_{0} \right\} \Tr \left\{ \mathcal{I}_{\alpha} \rho_{i} \right\}, \\
	c_{i}^{\alpha} &= \Tr \left\{ \mu_{i}^{\dagger} \rho_{0} \right\} \Tr \left\{ \mathcal{I}_{\alpha} \xi_{i} \right\}, \\
	d_{i j}^{\alpha} &= \Tr \left\{ \zeta_{i}^{\dagger} \rho_{0} \right\} \Tr \left\{ \mathcal{I}_{\alpha} \eta_{j} \right\}.
\end{align}
Following a similar algebraic procedure for the noise at steady state, we can express it to distinguish the role of the spectral components of the Lindbladian. We obtain:
\begin{align}
	S_{\alpha \alpha'}^{ss} \left( \tau \right)
	&= \delta_{\alpha \alpha'} \delta \left( \tau \right) \Tr \left\{ \mathcal{A}_{\alpha} \rho_{ss} \right\} \notag \\
	& \quad + \theta \left( \tau \right) \left[ \sum_{i = 2}^{2^{2 N} - 2 p - m} b_{i}^{\alpha \alpha'} e^{\lambda_{i} \tau} + 2 \sum_{i = 1}^{p} e^{\Re \left( \theta_{i} \right) \tau} \left( \Re \left( c_{i}^{\alpha \alpha'} \right) \cos \left[ \Im \left( \theta_{i} \right) \tau \right] \right. \right. \notag \\
	& \qquad - \left. \left. \Im \left( c_{i}^{\alpha \alpha'} \right) \sin \left[ \Im \left( \theta_{i} \right) \tau \right] \right) + e^{\gamma \tau} \sum_{i = 1}^{m} \sum_{j = 1}^{i} \frac{d_{i j}^{\alpha \alpha'}}{\left( i - j \right) !} \tau^{i - j} \right] \notag \\
	& \quad + \theta \left( -\tau \right) \left[ \sum_{i = 2}^{2^{2 N} - 2 p - m} b_{i}^{\alpha' \alpha} e^{-\lambda_{i} t} + 2 \sum_{i = 1}^{p} e^{-\Re \left( \theta_{i} \right) \tau} \left( \Re \left( c_{i}^{\alpha' \alpha} \right) \cos \left[ \Im \left( \theta_{i} \right) \tau \right] \right. \right. \notag \\
	& \qquad + \left. \left. \Im \left( c_{i}^{\alpha' \alpha} \right) \sin \left[ \Im \left( \theta_{i} \right) \tau \right] \right\} + e^{-\gamma \tau} \sum_{i = 1}^{m} \sum_{j = 1}^{i} \frac{d_{i j}^{\alpha' \alpha}}{\left( i - j \right) !} \left( -\tau \right)^{i - j} \right], \label{eq:noise_tau}
\end{align}
with 
\begin{align}
	b_{i}^{\alpha \alpha'}
    &= \Tr \left\{ \sigma_{i}^{\dagger} \mathcal{I}_{\alpha} \rho_{ss} \right\} \Tr \left\{ \mathcal{I}_{\alpha'} \rho_{i} \right\}, \\
	c_{i}^{\alpha \alpha'}
    &= \Tr \left\{ \mu_{i}^{\dagger} \mathcal{I}_{\alpha} \rho_{ss} \right\} \Tr \left\{ \mathcal{I}_{\alpha'} \xi_{i} \right\}, \\
	d_{i j}^{\alpha}
    &= \Tr \left\{ \zeta_{i}^{\dagger} \mathcal{I}_{\alpha} \rho_{ss} \right\} \Tr \left\{ \mathcal{I}_{\alpha'} \eta_{j} \right\}.
\end{align}
Eq.~\eqref{eq:currents} and Eq.~\eqref{eq:noise_tau} explicitly show the role of the spectral features of the Lindbladian into the dynamics of transport quantities. In particular, the following statements are valid in general:
\begin{itemize}
\item real and distinct eigenvalues lead to exponential decay corresponding to the overdamped regime;
\item complex eigenvalues lead to oscillations in the presence of exponential decay corresponding to the underdamped regime. Their real parts determine decaying rates and imaginary parts decide periods of oscillations;
\item the polynomial dependence is the result of real and identical eigenvalues that correspond to exceptional points. 
\end{itemize}
Remarkably, Eq.~\eqref{eq:noise_tau} demonstrates that the noise at steady state exhibits genuine features corresponding to exceptional points. This results opens the possibility to witness and exploit these non-Hermitian features at long times. We emphasize that the above eqautions are valid for arbitrary quantum systems coupled to multiple environments, assuming their dynamics being given by a Lindblad type master equation.

\subsection{Signatures of exceptional points in the noise spectrum}

Auto-correlation functions $S_{\alpha \alpha}^{ss}$ are symmetric with respect to $\tau = 0$, therefore they are continuous; while cross-correlation functions exhibit a discontinuity at $\tau = 0$
\begin{align}
	\lim_{\tau \to 0^{+}} S_{\alpha \alpha'}^{ss} \left( \tau \right) - \lim_{\tau \to 0^{-}} S_{\alpha \alpha'}^{ss} \left( \tau \right)
    &= \sum_{i = 2}^{2^{2 N} - 2 p - m} \left( b_{i}^{\alpha \alpha'}  - b_{i}^{\alpha' \alpha} \right) + 2 \sum_{i = 1}^{p} \Re \left( c_{i}^{\alpha \alpha'} - c_{i}^{\alpha' \alpha} \right) \notag \\
    & \quad + \sum_{i = 1}^{m} \left( d_{i i}^{\alpha \alpha'} - d_{i i}^{\alpha' \alpha} \right).
\end{align}
The Fourier transform of Eq.~\eqref{eq:noise_tau} reads
\begin{align}
	S_{\alpha \alpha'}^{ss} \left( \omega \right)
	&= \int_{-\infty}^{+\infty} S_{\alpha \alpha'}^{ss} \left( \tau \right) e^{i \omega \tau} d\tau \notag \\
	&= \delta_{\alpha \alpha'} \Tr \left\{ \mathcal{A}_{\alpha} \rho_{ss} \right\} - \sum_{i = 2}^{2^{2 N} - 2 p - m} \left( \frac{b_{i}^{\alpha \alpha'}}{\lambda_{i} + i \omega} + \frac{b_{i}^{\alpha' \alpha}}{\lambda_{i} - i \omega} \right)\notag \\
    & \quad - \sum_{i = 1}^{p} \left( \frac{c_{i}^{\alpha \alpha'}}{\theta_{i} + i \omega} + \frac{c_{i}^{\alpha' \alpha}}{\theta_{i} - i \omega} + \frac{{c'}_{i}^{\alpha \alpha'}}{\theta_{i}^{*} + i \omega} + \frac{{c'}_{i}^{\alpha' \alpha}}{\theta_{i}^{*} - i \omega} \right) \notag \\
	& \quad + \sum_{i = 1}^{m} \sum_{j = 1}^{i} \left( -1 \right)^{i - j + 1} \left( \frac{d_{i j}^{\alpha \alpha'}}{\left( \gamma + i \omega \right)^{i - j + 1}} + \frac{d_{i j}^{\alpha' \alpha}}{\left( \gamma - i \omega \right)^{i - j + 1}} \right). \label{eq:noise_f}
\end{align}

Equation~\eqref{eq:noise_f} is very informative. The eigenvalues of the Lindbladian $\mathcal{L}$ and their opposite values correspond exactly to the poles of the noise in the frequency domain \cite{arkhipov2020liouvillian}. 
A pole of order larger than one signals the presence of a non-trivial Jordan block and therefore provides a direct signature of a Lindbladian exceptional point. The maximal pole order determines the order of the corresponding exceptional point. We note that related signatures of exceptional points in the frequency-domain structure of correlation functions have recently been discussed in Ref.~\cite{molina2026spectroscopic}.
From Eq.~\eqref{eq:noise_f}, we also obtain exact expressions of shot noise which are discussed in Sec.~\ref{sec:two-qubit}.
\begin{align}
	S_{\alpha \alpha'}^{ss} \left( \omega = 0 \right)
    &= \delta_{\alpha \alpha'} \Tr \left\{ \mathcal{A}_{\alpha} \rho_{ss} \right\} - \sum_{i = 2}^{2^{2 N} - 2 p - m} \frac{b_{i}^{\alpha \alpha'} + b_{i}^{\alpha' \alpha}}{\lambda_{i}} - \sum_{i = 1}^{p} \left( \frac{c_{i}^{\alpha \alpha'} + c_{i}^{\alpha' \alpha}}{\theta_{i}} + \frac{{c'}_{i}^{\alpha \alpha'} + {c'}_{i}^{\alpha' \alpha}}{\theta_{i}^{*}} \right) \notag \\
    & \quad + \sum_{i = 1}^{m} \sum_{j = 1}^{i} \frac{d_{i j}^{\alpha \alpha'} + d_{i j}^{\alpha' \alpha}}{\left( -\gamma \right)^{i - j + 1}}. \label{eq:shot_noise}
\end{align}

In the next section, we consider a paradigmatic example of an open quantum system, for which the above quantities and signatures can be discussed based on analytical calculations. We consider a two-qubit system, made of two interacting qubits, each of them being weakly coupled to their orn environments. This model has been proven useful in the past for demonstrating the possibility of creating and manipulating entanglement through dissipation \cite{brask2015autonomous, khandelwal2024chiral}, as well as for discussing signatures of EPs in the dynamics of an open quantum system \cite{khandelwal2021signatures, khandelwal2025emergent}.

\section{Illustration: A two-qubit model} 
\label{sec:two-qubit}

\begin{figure}[h]
	\centering
 \includegraphics[width=0.45\linewidth]{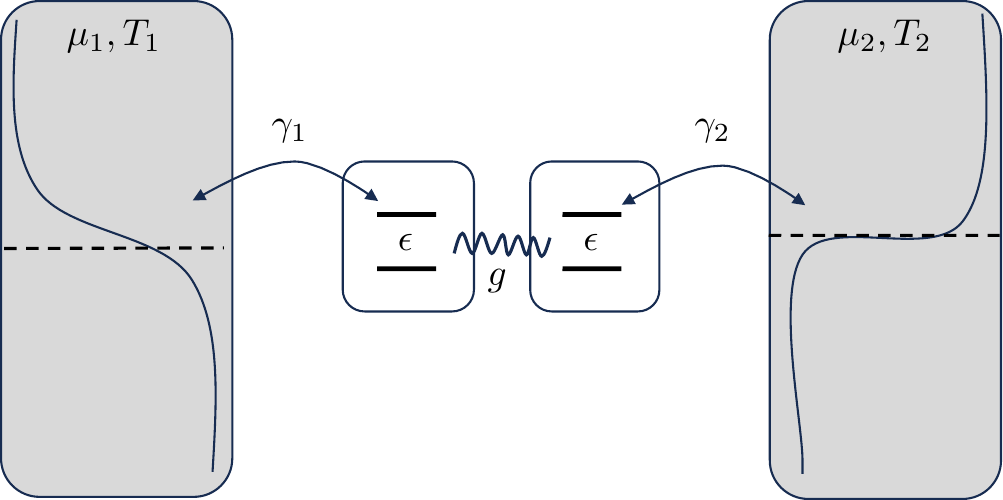}
		
	\caption{Sketch of a two-qubit system with the following parameters, bare energy of the two qubits $\epsilon$, the inter-qubit coupling strength $g$, reservoirs characterized by their respective chemical potential and temperature $\mu_i, T_i, i=1,2$ and bare bath-reservoir coupling strengths $\gamma_{1}$ and $\gamma_{2}$.}
	\label{fig:two-qubit}
\end{figure}

We now consider a weakly-coupled two-qubit model to illustrate our general results discussed in the previous section. The model is shown in Fig.~\ref{fig:two-qubit}: two identical interacting qubits with inter-coupling strength $g$ are weakly coupled to their own reservoir respectively. The qubits are energy-degenerate with a bare energy gap $\epsilon$ and the reservoir-system coupling strengths are $\gamma_{1}$ and $\gamma_{2}$. 

The Hamiltonian for the two interacting qubits follows the generic form of the Hamiltonian in Eq.~\eqref{eq:ham_syst} with $j=1,2$ and the two qubits being energy degenerate:
\begin{equation}
	H_{S} = \epsilon \left( \sigma_{+}^{\left( 1 \right)} \sigma_{-}^{\left( 1 \right)} + \sigma_{+}^{\left( 2 \right)} \sigma_{-}^{\left( 2 \right)} \right) + g \left( \sigma_{+}^{\left( 1 \right)} \sigma_{-}^{\left( 2 \right)} + \sigma_{-}^{\left( 1 \right)} \sigma_{+}^{\left( 2 \right)} \right)\,.
\end{equation}
Its dynamics is given by the Lindblad equation in Eq.~\eqref{eq:Lindbladian} if $\gamma_{i} \ll \epsilon$, $g \ll \epsilon$ and $g \lesssim \gamma_{i}$ to ensure its validity as discussed previously. For a two-qubit system, the dimension of the density matrix $\rho$ is $4 \times 4$ and hence, the Lindbladian $\mathcal{L}$ has a dimension of $16 \times 16$. Here we recall results found in \cite{khandelwal2021signatures}. To intentionally reduce the dimension of the Liouville space, we exploit the fact that the steady state of the two qubits only has six nonzero elements: the four populations of the four computational states and two coherences (off-diagonal terms) $\ket{01}\bra{10}$ and $\ket{10}\bra{01}$. All other elements decay exponentially to zero, without being affected by the unitary dynamics induced by $H_S$. If we consider initial states that belong to this steady-state subspace, the dynamics is fully captured within this $6 \times 6$ Liouville subspace. The matrix form of the reduced $L$, denoted $\tilde{L}$, assuming the basis $\{\ket{00}\bra{00}$, $\ket{01}\bra{01}$, $\ket{01}\bra{10}$, $\ket{10}\bra{01}$, $\ket{10}\bra{10}$, $\ket{11}\bra{11}\}$ takes the form:
\begin{equation}
	\tilde{L} = 
	\begin{pmatrix}
		-\gamma_{+} & \gamma_{2}^{-} & 0 & 0 & \gamma_{1}^{-} & 0 \\
		\gamma_{2}^{+} & -\gamma_{1}^{+} - \gamma_{2}^{-} & ig & -ig & 0 & \gamma_{1}^{-} \\
		0 & ig & -\frac{1}{2} \gamma & 0 & -ig & 0 \\
		0 & -ig & 0 & -\frac{1}{2} \gamma & ig & 0 \\
		\gamma_{1}^{+} & 0 & -ig & ig & -\gamma_{1}^{-} - \gamma_{2}^{+} & \gamma_{2}^{-} \\
		0 & \gamma_{1}^{+} & 0 & 0 & \gamma_{2}^{+} & -\gamma_{-} \label{eq:reduced_L}
	\end{pmatrix}
\end{equation}
with the simplified notations $\gamma_{-} = \gamma_{1}^{-} + \gamma_{2}^{-}$, $\gamma_{+} = \gamma_{1}^{+} + \gamma_{2}^{+}$ and $\gamma = \gamma_{-} + \gamma_{+}$. The in-tunneling rates $\gamma_{i}^{+}$ and out-tunneling rates $\gamma_{i}^{-}$ are the product of bare tunneling rates and the occupation probability of the reservoir $i$. For fermionic reservoirs, they are given by:
\begin{equation}
\gamma_\alpha^{+} = \gamma_\alpha n_F(\epsilon, T_\alpha) \quad ; \quad \gamma_\alpha^{-} = \gamma_\alpha (1-n_F(\epsilon, T_\alpha))\,. 
\end{equation}
and for bosonic reservoirs, they read:
\begin{equation}
\gamma_\alpha^{+} = \gamma_\alpha n_B(\epsilon, T_\alpha) \quad ; \quad \gamma_\alpha^{-} = \gamma_\alpha (1 +n_B(\epsilon, T_\alpha))\,,
\end{equation}
with $n_{F,B}(\epsilon, T_\alpha)$  the Fermi-Dirac or Bose-Einstein distribution evaluated at energy $\epsilon$ of the qubits and temperature $T_\alpha$ of the corresponding reservoir. The spectrum of $\tilde{L}$, denoted $\sigma(\tilde{L})$, can be computed analytically:
\begin{equation}
	\sigma (\tilde{L}) = \left\{ \lambda_{1} = 0, \lambda_{2} = -\gamma, \lambda_{3} = \lambda_{4} =-\frac{\gamma}{2}, \lambda_{5} = -\frac{\gamma}{2} + \eta, \lambda_{6} = -\frac{\gamma}{2} - \eta \right\}
\end{equation}
with $\eta = \sqrt{ \Delta\gamma^{2} - 4 g^{2}}$ and $\Delta\gamma = (\gamma_{1} - \gamma_{2})/2$ and $\gamma_{i} = \gamma_{i}^{-} + \gamma_{i}^{+}$. The corresponding left- and right- eigenmatrices are listed in the Appendix~\ref{sec:appendix}. Figure~\ref{fig:distribution} illustrates the spectrum $\sigma( \tilde{L})$ on a complex plane.

\begin{figure}
	\centering
	\begin{tikzpicture}
		\draw[line width = 0.8pt][->] (-7, 0)--(1, 0);
		\draw[line width = 0.8pt][->] (0, -3)--(0, 3);
		\draw[line width = 0.8pt] (-0.1, -0.1)--(0.1, 0.1);
		\draw[line width = 0.8pt] (-0.1, 0.1)--(0.1, -0.1);
		\draw[line width = 0.8pt] (-6.1, -0.1)--(-5.9, 0.1);
		\draw[line width = 0.8pt] (-6.1, 0.1)--(-5.9, -0.1);
		\draw[line width = 0.8pt] (-3.1, -0.1)--(-2.9, 0.1);
		\draw[line width = 0.8pt] (-3.1, 0.1)--(-2.9, -0.1);
		\draw[fill = red] (-3, 1.5) circle (0.1);
		\draw[fill = red] (-3, -1.5) circle (0.1);
		\draw[line width = 0.8pt][draw = red][->] (-3, 1.2)--(-3, 0.6);
		\draw[line width = 0.8pt][draw = red][->] (-3, -1.2)--(-3, -0.6);
		\draw[fill = blue] (-5, 0) circle (0.1);
		\draw[fill = blue] (-1, 0) circle (0.1);
		\draw[line width = 0.8pt][draw = blue][->] (-4.8, 0.2)--(-4.2, 0.2);
		\draw[line width = 0.8pt][draw = blue][->] (-1.2, 0.2)--(-1.8, 0.2);

		\node at (1.5, 0) {$\Re \left( \lambda_{i} \right)$};
		\node at (0.6, 3) {$\Im \left( \lambda_{i} \right)$};
		\node at (0.2, -0.4) {$0$};
		\node at (0.2, 0.4) {$\lambda_{1}$};
		\node at (-6, -0.4) {$-\gamma$};
		\node at (-6, 0.4) {$\lambda_{2}$};
		\node at (-3, -0.4) {$-\frac{\gamma}{2}$};
		\node at (-3.5, 0.4) {$\lambda_{3}$};
		\node at (-2.5, 0.4) {$\lambda_{4}$};
		\node at (-3, 1.9) {$\lambda_{5}$};
		\node at (-2, 1.5) {$-\frac{\gamma}{2} + i \eta$};
		\node at (-3, -1.9) {$\lambda_{6}$};
		\node at (-2, -1.5) {$-\frac{\gamma}{2} - i \eta$};
		\node at (-1, 0.4) {$\lambda_{5}$};
		\node at (-1.2, -0.4) {$-\frac{\gamma}{2} + \eta$};
		\node at (-5, 0.4) {$\lambda_{6}$};
		\node at (-4.8, -0.4) {$-\frac{\gamma}{2} - \eta$};
	\end{tikzpicture}
	\caption{Spectrum of the reduced Lindbladian $\sigma \left( \tilde{L} \right)$ on a complex plane showing the coalescence of $\lambda_{4}$, $\lambda_{5}$ and $\lambda_{6}$ at a third-order exceptional point. Color blue represents overdamped regime where $\eta$ is real while red meas underdamped regime and we change $\eta$ to $i \eta$.}
	\label{fig:distribution}
\end{figure}
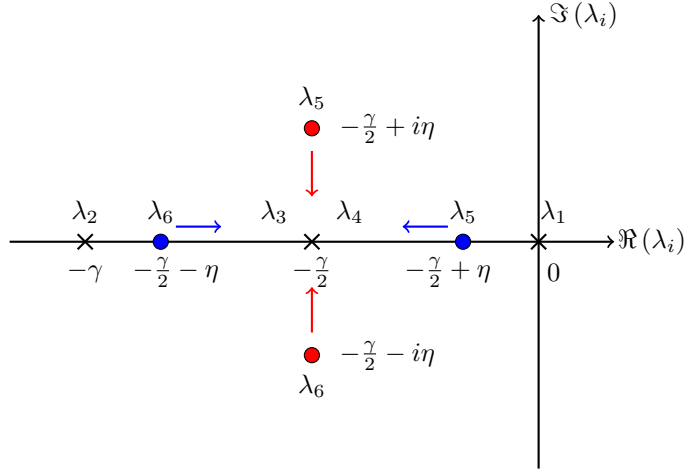

For this model, it is known that there exists a unique steady-state corresponding to $\lambda_{1} = 0$. The parameter $\eta$, which depends on the set of parameters $\xi = \left\{ \epsilon, g, \gamma_{1}, \gamma_{2}, T_{1}, T_{2} \right\}$, determines the existence of exceptional points.

\begin{itemize}

\item If $\eta=0$, corresponding to $g = \left| \gamma_{1} - \gamma_{2} \right|/4$, a third-order exceptional point arises, corresponding to a critical regime.
\item If $\eta^{2} > 0$, $\sigma(\tilde{L})$ lies on the real plane, corresponding to the overdamped regime. We expect exponentially decaying behavior of transport observables. 
\item If $\eta^{2}<0$, corresponding to $g > \left| \gamma_{1} - \gamma_{2} \right|/4$. the eigenvalues $\lambda_{5}$ and $\lambda_{6}$ become complex, corresponding to the underdamped regime. We should observe oscillations in the presence of decay for transport observables. 
\end{itemize}
Moreover, we note that $\Re \left( \lambda_{i} \right) < 0$ whatever $\xi$ is, except for $\lambda_{1} = 0$.

\subsection{Steady-state currents}

In the long time limit, the expressions of the currents in both reservoirs are known from previous literature and straightforward to derive, see Refs. \cite{khandelwal2020critical, blasi2024exact} and references therein. We recall them below for clarity in the manuscript. 
\begin{equation}
	I_{1}^{ss} = -I_{2}^{ss} = \frac{4g^{2} \left( \gamma_{1}^{+} \gamma_{2}^{-} - \gamma_{1}^{-} \gamma_{2}^{+} \right)}{\left( 4g^{2} + \gamma_{1} \gamma_{2} \right) \gamma} = \frac{4 g^{2}}{4 g^{2} + \gamma_{1} \gamma_{2}} \frac{\gamma_{1} \gamma_{2}}{\gamma} (n_1 - n_2) =  \frac{4 g^{2}}{\mathcal{G}_0} \frac{\gamma_{1} \gamma_{2}}{\gamma} (n_1 - n_2)\,.\label{eq:currents_steady}
\end{equation}
We have introduced the notation $\mathcal{G}_0 = 4 g^2 + \gamma_1 \gamma_2$ as this factor in the denominator will regularly appear in subsequent derivations. The opposite sign between $I_1^{ss}$ and $I_2^{ss}$ reflects current conservation at steady state. In contrast, at finite times, $I_{1} \left( t \right) \neq I_{2} \left( t \right)$ \cite{bourgeois2024finite}.
The factor $\gamma_{1} \gamma_{2}/\gamma$ corresponds to the probability of particle exchange between reservoirs $1$ and $2$, while the factor $4 g^2/(4 g^{2} + \gamma_{1} \gamma_{2})$ modifies this probability accounting for the qubit interaction. Here $n_1, n_2$ are the Bose-Einstein or Fermi-Dirac distributions, setting the transport window. It is clear from these expressions and the discussion in the previous section that no signature of EPs can be witnessed in these average steady-state currents. In contrast, their expressions at finite times, i.e. in the transient regime, do exhibit signatures of EPs as discussed below.\\


\subsection{Transient currents}
\label{transient-currents}

We now provide the full expressions of the currents $I_{1}$ and $I_{2}$ at
finite times as a function of the parameter $\eta$, in a form that unifies
the three dynamical regimes.
To this end, we introduce the functions
\begin{equation}
\Phi_C(x)=
\begin{cases}
\cosh x, & \eta^2>0,\\
\cos x, & \eta^2<0,
\end{cases}
\qquad
\Phi_S(x)=
\begin{cases}
\sinh x, & \eta^2>0,\\
i \sin x, & \eta^2<0,
\end{cases}
\label{eq:Phi_def}
\end{equation}
which allow us to treat overdamped and underdamped dynamics with a unified and simplified notation. We keep the $\eta$-dependence explicit as the critical regime will follow as the regular limit $\eta\to 0$.

\begin{figure*}
	\centering
	\subcaptionbox{
		A plot of $I_{1} \left( t \right)$ in three different regimes \label{fig:current1}
	}{
		\includegraphics[width=0.45\linewidth]{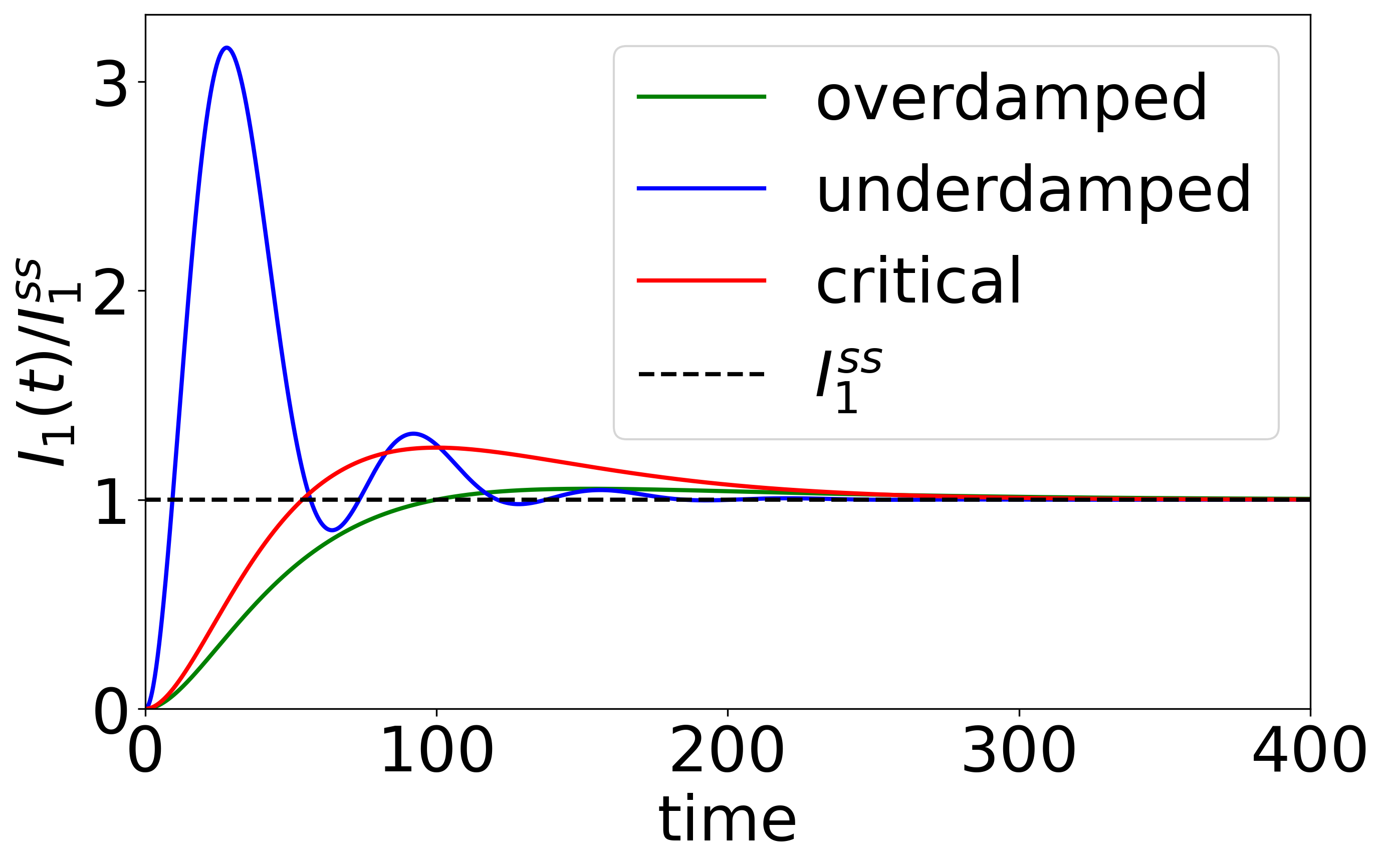}
	}
	\hfill
	\subcaptionbox{
		A plot of $I_{2} \left( t \right)$ in three different regimes \label{fig:current2}
	}{
		\includegraphics[width=0.45\linewidth]{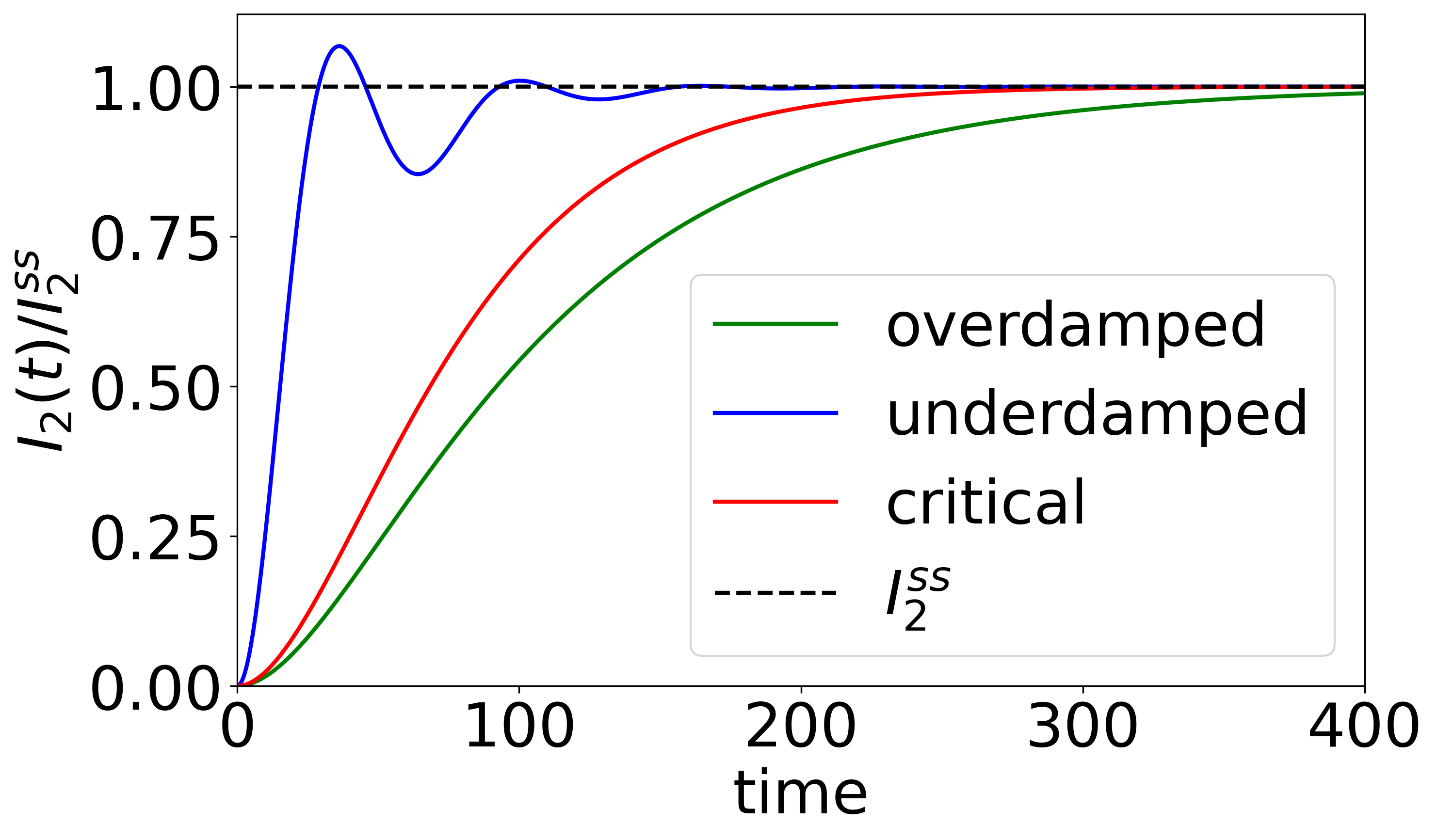}
	}
	\caption{Currents as a function of time. \ref{fig:current1} and \ref{fig:current2} show the dynamics of $I_{1} \left( t \right)$ and $I_{2} \left( t \right)$ respectively. Both are normalized by their steady-state values. The initial steady is a product of thermal state. We choose the following set of parameters $\xi = \left\{ \epsilon = 1, \gamma_{1} = 0.05, \gamma_{2} = 0.01, T_{1} = 1, T_{2} = 0.5 \right\}$. And the coupling strength $g$ in each regime are $g_{\eta^2>0} = \bar{g}/2$ and $g_{\eta^2<0} = 5 \bar{g}$, where $\bar{g} =\left| \gamma_{1} - \gamma_{2} \right|/4 = 0.01$ represents the value of $g$ at exceptional point. In \ref{fig:current1}, two inflexion points occur in both overdamped and ciritical regimes for current $1$, while both $I_{2}^{overdamped}$ and $I_{2}^{critical}$ monotonically grow as \ref{fig:current2} shows.}
	\label{fig:currents}
\end{figure*}

For $\eta\neq 0$, the normalized currents can be written as
\begin{equation}
\label{eq:currents_transient}
    \frac{I_{\alpha}^{\eta \neq 0} \left( t \right)}{I_{\alpha}^{ss}} = 1 + e^{-\frac{\gamma}{2} t} \left[ \mathcal{J}_{\alpha, 0} + \mathcal{J}_{\alpha, C} \Phi_{C} \left( \eta t \right) + \mathcal{J}_{\alpha, S} \Phi_{S} \left( \eta t \right) \right],
\end{equation}
where
\begin{align}
    \mathcal{J}_{\alpha, 0}
    &= \frac{\frac{\gamma}{2} - \gamma_{\alpha}}{\left( \gamma - \gamma_{\alpha} \right) \eta^{2}} \mathcal{G}_{0}, \\
    \mathcal{J}_{\alpha, C}
    &= \frac{\left[ 4 g^{2} + \left( \gamma - \gamma_{\alpha} \right) \left( \gamma_{\alpha} - \frac{\gamma}{2} \right) \right] \gamma}{2 \left( \gamma - \gamma_{\alpha} \right) \eta^{2}}, \\
    \mathcal{J}_{\alpha, S}
    &= -\frac{\gamma}{2 \eta}.
\end{align}
We recall that
\begin{equation}
    \mathcal{G}_0 \equiv 4g^2+\gamma_1\gamma_2\,.
\end{equation}

For $\eta^2>0$, Eq.~\eqref{eq:currents_transient} yields hyperbolic functions and the
dynamics is overdamped. For $\eta^2<0$, the underdamped regime is obtained
by the analytic continuation $\eta\to i\eta$, which maps $\cosh$ and $\sinh$
into $\cos$ and $\sin$, leading to oscillating currents. 
We note that non-monotonic relaxation may still occur in the overdamped regime, including transient extrema, due to the competition between distinct real decay modes. This should not be confused with underdamped oscillations, which instead originate from complex-conjugate Lindbladian eigenvalues. The critical regime follows from the regular limit $\eta\to 0$. Using
\begin{equation}
\Phi_C(\eta t)=1+\frac{\eta^2 t^2}{2}+\mathcal O(\eta^4),
\qquad
\Phi_S(\eta t)=\eta t+\mathcal O(\eta^3),
\end{equation}
one finds that the apparent divergences cancel out, leading to
\begin{equation}
    \frac{I_{\alpha}^{\eta = 0}}{I_{\alpha}^{ss}} = 1 + e^{-\frac{\gamma}{2}} \left( -1 - \frac{\gamma}{2} t + \frac{\left( \gamma_{\alpha} - \frac{\gamma}{2} \right) \gamma^{2}}{8 \left( \gamma - \gamma_{\alpha} \right)} t^{2} \right).
\end{equation}
In the critical regime, oscillations disappear, and a time-polynomial behaviour at short times is present, constituting a signature of exceptional points in averaged-like quantities as discussed in previous literature \cite{khandelwal2021signatures, zhou2023accelerating}\\

Figure~\ref{fig:currents} illustrates the distinct dynamical behaviors of the
system. Both $I_{1}$ and $I_{2}$ exhibit identical qualitative evolution.
In the overdamped regime, the currents approach their steady-state values
monotonically, possibly with transient overshoots. In the underdamped regime,
the approach to equilibrium is accompanied by damped oscillations. The
critical regime lies at the boundary between these two behaviors, separating
purely exponential relaxation from oscillatory dynamics.

\subsection{\texorpdfstring{$\tau$}{tau}-dependent steady-state noise}
\label{tau-Noise}
We also obtain explicit expressions of time-dependent steady-state noises, which can be written in unified and compact forms.
We recall the total tunneling rates
\begin{equation}
\gamma_\pm \equiv \gamma_1^\pm+\gamma_2^\pm,
\qquad
\gamma \equiv \gamma_+ + \gamma_-,
\end{equation}
and introduce the compact combinations
\begin{align}
    \mathcal{G}_{0}
    &= 4 g^{2} + \gamma_{1} \gamma_{2}, \\
    \mathcal{G}_{\alpha, 1}&
    = 4 g^{2} \left( \gamma_{\alpha}^{-} \gamma_{+}^{2} + \gamma_{\alpha}^{+} \gamma_{-}^{2} \right) + \gamma_{\alpha}^{-} \gamma_{\alpha}^{+} \left( \gamma - \gamma_{\alpha} \right) \gamma^{2}, \\
    \mathcal{G}_{\alpha, 2}&
    = 4 g^{2} \left[ \gamma_{\alpha}^{-} \gamma_{\alpha}^{+} \gamma^{2} - \left( \gamma_{1}^{-} \gamma_{2}^{+} - \gamma_{1}^{+} \gamma_{2}^{-} \right)^{2} \right] + \gamma_{\alpha}^{-} \gamma_{\alpha}^{+} \gamma_{1} \gamma_{2} \gamma^{2}, \\
    \mathcal{F}_{\alpha}&
    = 4 g^{2} + \left( -1 \right)^{\alpha} \gamma_{\alpha} \Delta\gamma.
\end{align}
The singular contribution reads
\begin{equation}
    \mathcal{S}_{\alpha, \delta} = \frac{4 g^{2} \left( \gamma_{\alpha}^{-} \gamma_{+} + \gamma_{\alpha}^{+} \gamma_{-} \right) + 2 \gamma_{\alpha}^{-} \gamma_{\alpha}^{+} \left( \gamma - \gamma_{\alpha} \right) \gamma}{\gamma \mathcal{G}_{0}}.
\end{equation}
For $\eta\neq 0$, the steady-state noise admits the unified representation
\begin{align}
    S_{\alpha \alpha', \eta \neq 0}^{ss} \left( \tau \right) 
    &= \delta_{\alpha \alpha'} \delta \left( \tau \right) \mathcal{S}_{\alpha, \delta} + \theta \left( \tau \right) e^{-\frac{\gamma}{2} \tau} \left[ \mathcal{S}_{\alpha \alpha', 0} + \mathcal{S}_{\alpha \alpha', C} \Phi_{C} \left( \eta \tau \right) + \mathcal{S}_{\alpha \alpha', S} \Phi_{S} \left( \eta \tau \right) \right] \\
    & \quad + \theta \left( -\tau \right) e^{\frac{\gamma}{2} \tau} \left[ \mathcal{S}_{\alpha' \alpha, 0} + \mathcal{S}_{\alpha' \alpha, C} \Phi_{C} \left( \eta \tau \right) - \mathcal{S}_{\alpha' \alpha, S} \Phi_{S} \left( \eta \tau \right) \right], \label{eq:noise_unified_clean}
\end{align}
where $\Phi_C$ and $\Phi_S$ are defined in Eq.~\eqref{eq:Phi_def}.
The coefficients take the form

\begin{align}
   &\mathcal{S}_{\alpha \alpha', 0}= \frac{2 g^{2} \gamma_{\alpha'}}{\gamma^{2} \eta^{2} \mathcal{G}_{0}} \mathcal{G}_{\alpha, 1}, \\
    &\mathcal{S}_{\alpha \alpha', C}= \left( -1 \right)^{\alpha - \alpha'} \frac{\gamma_{\alpha'}}{2 \gamma^{2} \eta^{2} \mathcal{G}_{0}^{2}} \left[ \mathcal{F}_{\alpha} \left( \mathcal{F}_{\alpha'} - \left( -1 \right)^{\alpha'} \gamma \Delta\gamma \right) \mathcal{G}_{\alpha, 1} - \left( \gamma - \gamma_{\alpha'} \right) \eta^{2} \mathcal{G}_{\alpha, 2} \right], \\
    &\mathcal{S}_{\alpha \alpha', S}= \left( -1 \right)^{\alpha - \alpha'} \frac{\gamma_{\alpha'}}{2 \gamma^{2} \eta \mathcal{G}_{0}^{2}} \left[ -\left( \gamma - \gamma_{\alpha'} \right) \mathcal{F}_{\alpha} \mathcal{G}_{\alpha, 1} + \left( \mathcal{F}_{\alpha'} - \left( -1 \right)^{\alpha'} \gamma \Delta\gamma \right) \mathcal{G}_{\alpha, 2} \right].
\end{align}
The three dynamical regimes are obtained as in the case of the currents:
for $\eta^2>0$ the decay is overdamped, while for $\eta^2<0$ the analytic
continuation $\eta\to i\eta$ yields underdamped oscillations.
The critical regime follows from the regular limit $\eta\to 0$, using the
same expansions of $\Phi_C$ and $\Phi_S$ as discussed in the previous
subsection. \\

\begin{figure*}
	\centering
	\subcaptionbox{
		A plot of $S_{11}^{ss} \left( \tau \right)$ in three different regimes \label{fig:noise11}
	}{
		\includegraphics[width=0.45\linewidth]{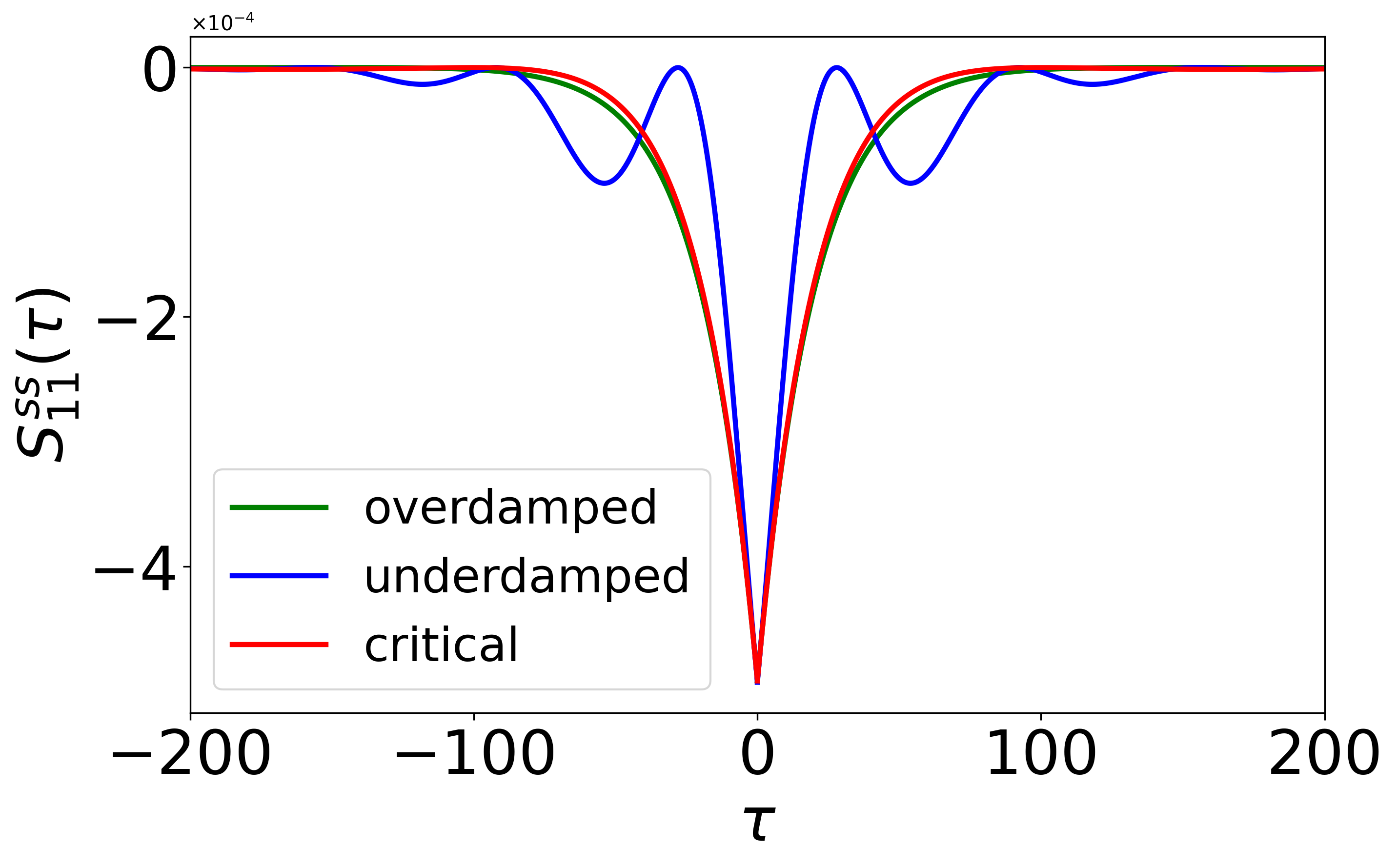}
	}
	\hfill
	\subcaptionbox{
		A plot of $S_{12}^{ss} \left( \tau \right)$ in three different regimes \label{fig:noise12}
	}{
		\includegraphics[width=0.45\linewidth]{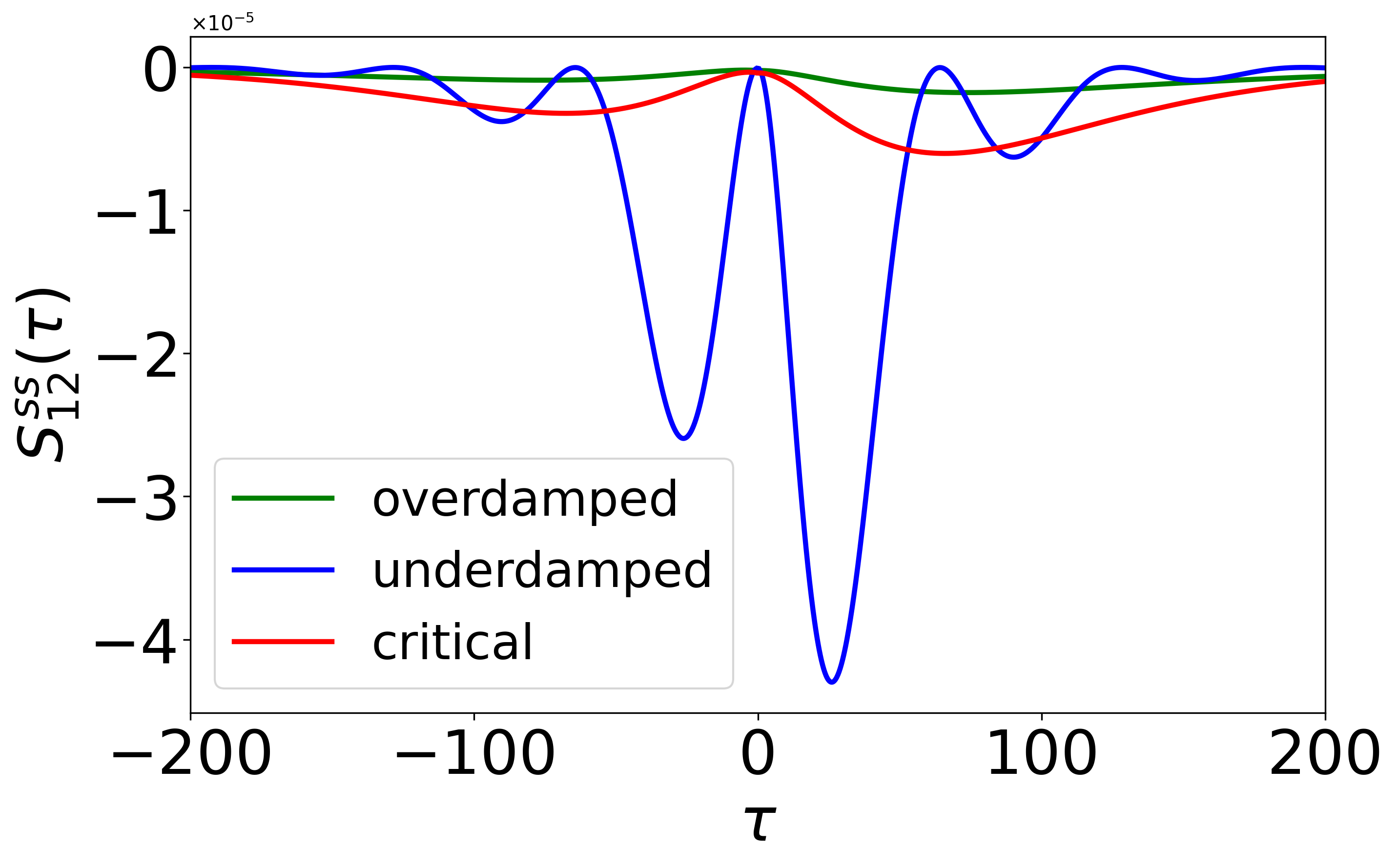}
	}
	\subcaptionbox{
		A plot of $S_{21}^{ss} \left( \tau \right)$ in three different regimes \label{fig:noise21}
	}{
		\includegraphics[width=0.45\linewidth]{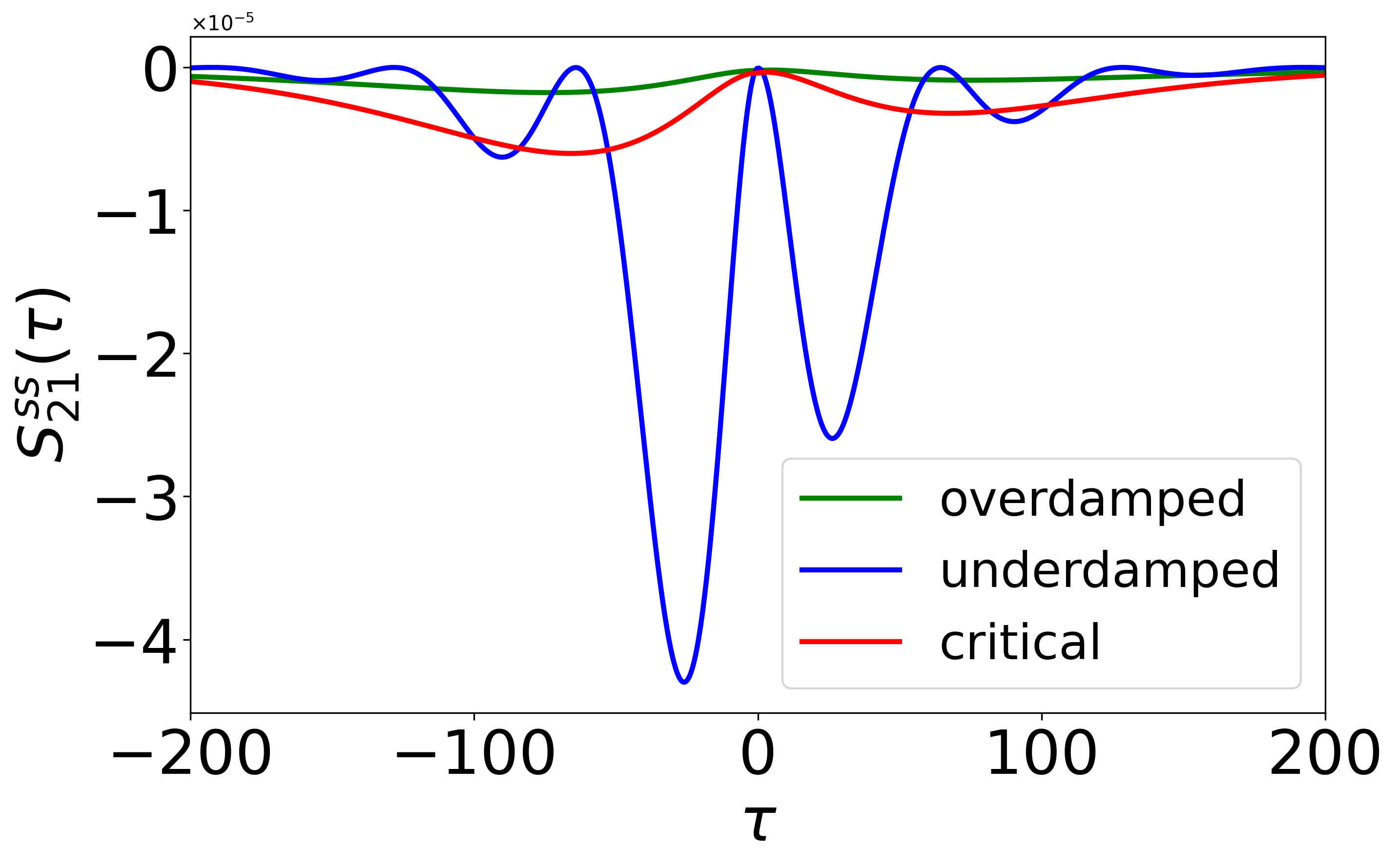}
	}
	\hfill
	\subcaptionbox{
		A plot of $S_{22}^{ss} \left( \tau \right)$ in three different regimes \label{fig:noise22}
	}{
		\includegraphics[width=0.45\linewidth]{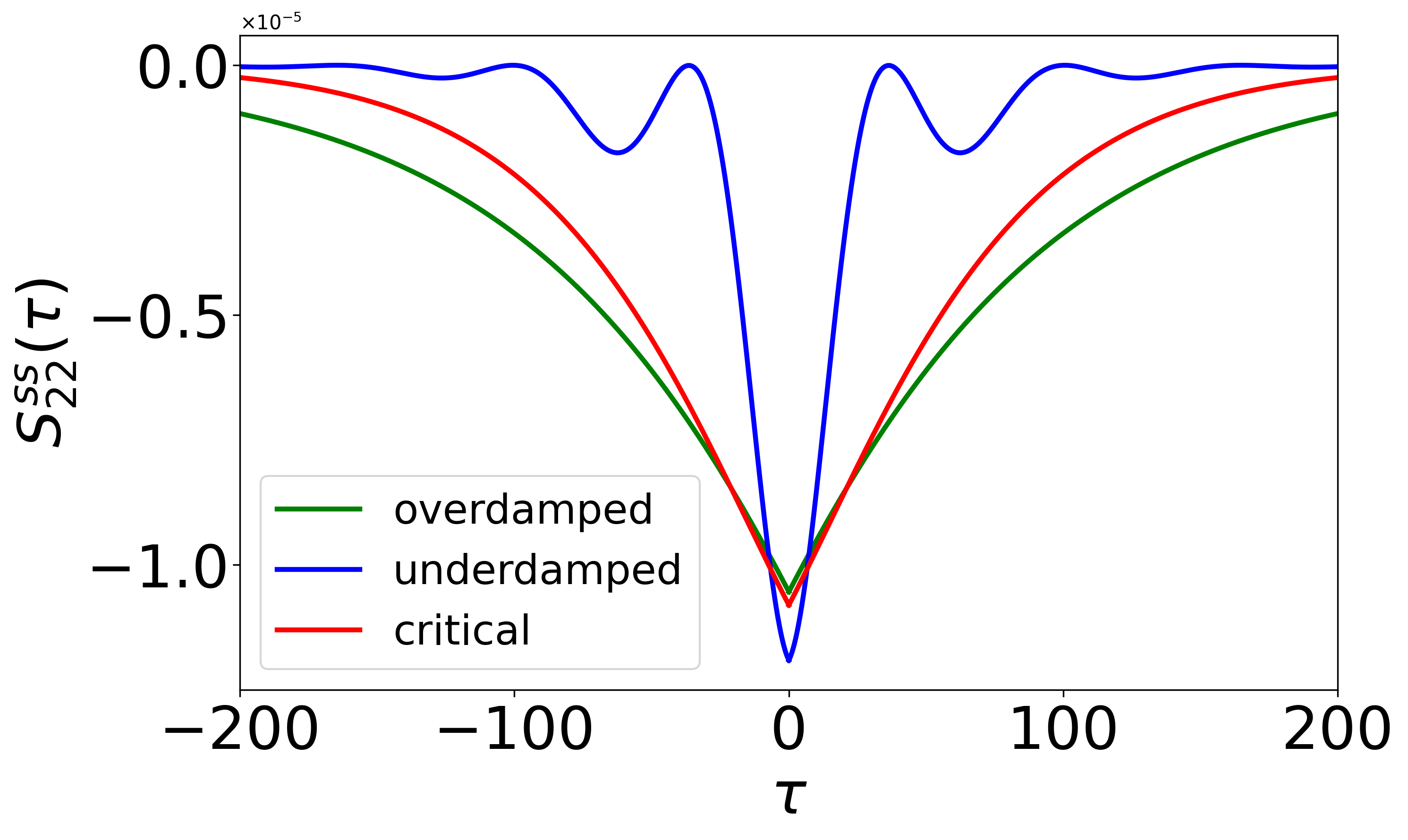}
	}
	\caption{
    Steady-state noise as a function of the time delay $\tau$. The four panels show the dynamics of $S_{11}^{ss} \left( \tau \right)$, $S_{12}^{ss} \left( \tau \right)$, $S_{21}^{ss} \left( \tau \right)$, and $S_{22}^{ss} \left( \tau \right)$.  The delta function $\delta(\tau)$ is treated as Kronecker delta $\delta_{\tau, 0}$.  We choose the following set of parameters $\xi = \left\{ \epsilon = 1, \gamma_{1} = 0.05, \gamma_{2} = 0.01, T_{1} = 1, T_{2} = 0.5 \right\}$. And the coupling strength $g$ in each regime are $g_{overdamped} = \frac{\bar{g}}{2}$ and $g_{underdamped} = 5 \bar{g}$, where $\bar{g} = \frac{\left| \gamma_{1} - \gamma_{2} \right|}{4} = 0.01$ represents the value of $g$ at exceptional point.}
	\label{fig:noise}
\end{figure*}

At the exceptional point defined by $\eta=0$, we obtain the following analytical forms for the steady-state noise functions:
\begin{align}
    S_{\alpha \alpha', \eta = 0}^{ss} \left( \tau \right)
    &= \delta_{\alpha \alpha'} \delta \left( \tau \right) \mathcal{S}_{\alpha, \delta} + \theta \left( \tau \right) e^{-\frac{\gamma}{2} \tau} \left( \mathcal{S}_{\alpha \alpha'}^{const} + \mathcal{S}_{\alpha \alpha'}^{lin} \tau + \mathcal{S}_{\alpha \alpha'}^{quad} \tau^{2} \right) \\
    & \quad + \theta \left( -\tau \right) e^{\frac{\gamma}{2} \tau} \left( \mathcal{S}_{\alpha' \alpha}^{const} - \mathcal{S}_{\alpha' \alpha}^{lin} \tau + \mathcal{S}_{\alpha' \alpha}^{quad} \tau^{2} \right)
\end{align}
with
\begin{align}
    &\mathcal{S}_{\alpha \alpha'}^{const}= \frac{\gamma_{\alpha'}}{4 \gamma^{2} \mathcal{G}_{0}^{2}} \left\{ \left[ \left( \alpha - \alpha' - 1 \right) \gamma_{1}^{2} - \left( \alpha - \alpha' + 1 \right) \gamma_{2}^{2} \right] \mathcal{G}_{\alpha, 1} - \left( -1 \right)^{\alpha - \alpha'} 2 \left( \gamma - \gamma_{\alpha'} \right) \mathcal{G}_{\alpha, 2} \right\}, \\
    &\mathcal{S}_{\alpha \alpha'}^{lin} = -\frac{\gamma_{\alpha'} \Delta\gamma}{4 \gamma \mathcal{G}_{0}^{2}} \left[ \left( -1 \right)^{\alpha'} \left( \gamma - \gamma_{\alpha'} \right) \mathcal{G}_{\alpha, 1} + \left( -1 \right)^{\alpha} \mathcal{G}_{\alpha, 2} \right], \\
    &\mathcal{S}_{\alpha \alpha'}^{quad}= -\frac{\gamma_{\alpha'} \Delta\gamma^{2}}{16 \mathcal{G}_{0}^{2}} \mathcal{G}_{\alpha, 1}.
\end{align}
Replacing the indices $\alpha, \alpha'$ by the reservoirs indices $\alpha, \alpha'=1,2$, we obtain the expresions of the auto- and cross-correlations functions of three different regimes shown in Fig.~\ref{fig:noise}.

This shows that the exceptional point manifests as a qualitative change in
the temporal structure of the noise, where oscillatory or hyperbolic
behavior collapses into a polynomial dependence on $|\tau|$.\\

Figure~\ref{fig:noise} illustrate the distinct behaviors corresponding to the underdamped, overdamped and critical regimes, for the different current correlation functions at steady state, as a function of the time delay $\tau$. As discussed based on our analytical results,  in the overdamped regimes, the noise functions are a sum of exponentials and therefore decay monotonically without any oscillations; in the underdamped regime, clear damped oscillations appear, with frequency $\eta$ and decay rate $\frac{\gamma}{2}$. At EP ($\eta=0$), the noise functions take the characteristic polynomial time-dependent form, hallmark of EPs. This illustrates the key finding of this work. By measuring second-order correlation functions of observables, it becomes possible to witness signatures of non-Hermitian EPs at long times. This has to be contrasted to signatures of EPs in averaged observables, only present at very short times (see discussion of the transient currents in the previous section). We believe that this result is very relevant towards experimental confirmation of EPs in nanoscale quantum devices. \\

\subsection{Signatures of EPs in the shot noise}
\label{shot-noise}
\begin{figure*}
	\centering
	\includegraphics[width=0.5\linewidth]{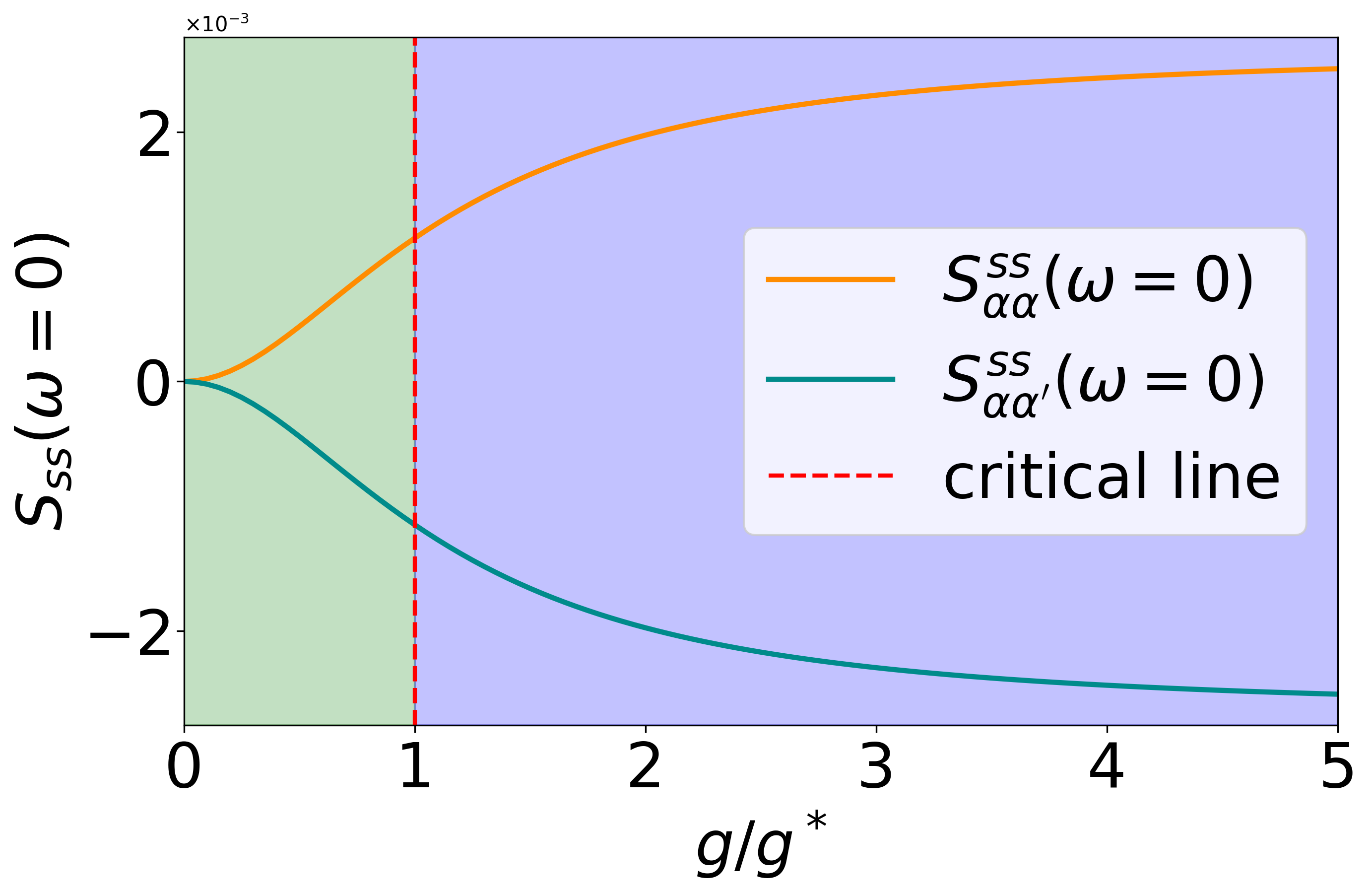}
	\caption{Shot noise as a function of $g$. Orange solid line and cyan solid represent shot noises of auto- and cross-correlation functions. Red dashed line corresponds to the critical line, which separates overdamped (light green area) and underdamped (light blue area) regimes.  We choose the following set of parameters $\xi = \left\{ \epsilon = 1, \gamma_{1} = 0.05, \gamma_{2} = 0.01, T_{1} = 1, T_{2} = 0.5 \right\}$. The range of $g$ is from $0$ to $5 g^*$, where $g^* = \frac{\left| \gamma_{1} - \gamma_{2} \right| }{4} = 0.01$ represents the value of $g$ at the exceptional point.}
	\label{fig:shot_noise}
\end{figure*}

We conclude this section by illustrating Eq.~\eqref{eq:shot_noise} with this specific model of two-interacting qubits. The zero-frequency noise is obtained by integrating
Eq.~\eqref{eq:noise_unified_clean} over the time delay,
\begin{equation}
    S_{\alpha \alpha', \eta \neq 0}^{ss} \left( \omega = 0 \right) = \delta_{\alpha \alpha'} \mathcal{S}_{\alpha, \delta} + \frac{2 \left( \mathcal{S}_{\alpha \alpha', 0} + \mathcal{S}_{\alpha' \alpha, 0} \right)}{\gamma} + \frac{\gamma \left( \mathcal{S}_{\alpha \alpha', C} + \mathcal{S}_{\alpha' \alpha, C} \right)}{2 \mathcal{G}_{0}} + \frac{\eta \left( \mathcal{S}_{\alpha \alpha', S} + \mathcal{S}_{\alpha' \alpha,S} \right)}{\mathcal{G}_{0}},
\end{equation}
while the critical expression follows from the regular limit $\eta\to 0$. Substituting the explicit expressions and using $\eta^2=\Delta\gamma^2-4g^2$, we obtain a symmetric form of shot noise, which shows explicit dependence on $g$
\begin{equation}
    S_{\alpha \alpha'}^{ss} \left( \omega = 0 \right) = \left( -1 \right)^{\alpha - \alpha'} \frac{4 g^{2}}{\gamma^{3} \mathcal{G}_{0}^{3}} \left( \mathcal{K}_{0} + 8 \mathcal{K}_{2} g^{2} + 16 \mathcal{K}_{4} g^{4} \right)
\end{equation}
with
\begin{align}
    \mathcal{K}_{0} &= \left( \gamma_{1}^{-} \gamma_{2}^{+} + \gamma_{1}^{+} \gamma_{2}^{-} \right) \gamma_{1}^{2} \gamma_{2}^{2} \gamma^{2}, \\
    \mathcal{K}_{2} &= \gamma_{1}^{-} \gamma_{1}^{+} \gamma_{2}^{4} + \gamma_{2}^{-} \gamma_{2}^{+} \gamma_{1}^{4} + 3 \left( \gamma_{1}^{-} \gamma_{1}^{+} \gamma_{2}^{2} + \gamma_{2}^{-} \gamma_{2}^{+} \gamma_{1}^{2} \right) \gamma_{1} \gamma_{2} + \left( \gamma_{1}^{-} - \gamma_{2}^{-} \right) \left( \gamma_{1}^{+} - \gamma_{2}^{+} \right) \gamma_{1}^{2} \gamma_{2}^{2}, \\
    \mathcal{K}_{4} &= \gamma^{2} \left( \gamma_{1}^{-} \gamma_{2}^{+} + \gamma_{1}^{+} \gamma_{2}^{-} \right) - 2 \left( \gamma_{1}^{-} \gamma_{2}^{+} - \gamma_{1}^{+} \gamma_{2}^{-} \right)^{2}.
\end{align}
Figure~\ref{fig:shot_noise} shows the shot noise as a function of
$g/g^*$, where $g^*$ denotes the critical coupling corresponding to the
exceptional point. As the system is tuned across this point, the shot noise
remains a smooth function, with no discontinuities or cusps. Hence, EP does not have a signature in the zero-frequency component of the spectral function of current correlations. However, as demonstrated in the previous section, the analytic properties of the finite-frequency expression  provide information about the order of existing EPs.

\section{Conclusion and Perspectives} \label{sec:ccl}
In this work, we have systematically investigated the influences imposed by Lindbladian exceptional points on dynamics of transport observables in open quantum systems, with a particular focus on steady-state regimes. By establishing a general theoretical framework based on Lindblad master equation and utilizing the spectral decomposition of the Lindbladian, we find that although the steady-state average currents are determined solely by the steady-state density matrix and thus cannot directly reflect non-Hermitian characteristics, higher-order current correlation functions, \textit{i.e.}, steady-state noise functions, can clearly exhibit signatures of LEPs. 
We illustrated these general results with the paradigmatic example of two weakly interacting qubits. By exactly solving the reduced Lindbladian, we provided analytical expressions for the finite-time currents and steady-state noise across the different dynamical regimes: overdamped, underdamped, and critical. Our analysis reveals that both the currents and the noise functions exhibit distinct behaviors in each regime.

 The theoretical framework developed here is applicable to arbitrary quantum systems coupled to multiple Markovian environments, suggesting that steady-state noise could serve as a general tool for detecting LEPs in a wide range of open quantum systems. As such, these results open several paths for future research and experimental exploration of non-Hermitian exceptional points at the quantum scale. 

 \section{Acknowledgements}

 G.H. acknowledges financial support during the duration of this project from NCCR SwissMAP and from the Sandoz Fundation through the FPFS-Monique de Meuron program for academic promotion. G.B. acknowledges Grant PID2023-151975NB-I00 funded by MICIU/AEI/10.13039/501100011033 and ERDF/EU.

\printbibliography

\section{Appendix} \label{sec:appendix}

\subsection{Properties of Lindbladian}
In this part, we are going to show that the eigenvalues of the Lindbladian $\mathcal{L}$ are either real or come in complex conjugate. Suppose $\lambda$ is one of the eigenvalues of $\mathcal{L}$ with the corresponding eigenmatrix $\rho$, satisfying $\mathcal{L} \rho = \lambda \rho$. Using the explicit form of $\mathcal{L}$ shown in \eqref{eq:Lindbladian}, we investigate the action of $\mathcal{L}$ on $\rho^{\dagger}$:
\begin{align}
    \mathcal{L} \rho^{\dagger}
    &= -i \left( H_{s} \rho^{\dagger} - \rho^{\dagger} H_{s} \right) + \sum_{\alpha} \sum_{j = 1}^{N} \left[ \gamma_{j \alpha}^{+} \left( \sigma_{+}^{\left( j \right)} \rho^{\dagger} \sigma_{-}^{\left( j \right)} - \frac{1}{2} \sigma_{-}^{\left( j \right)} \sigma_{+}^{\left( j \right)} \rho^{\dagger} - \frac{1}{2} \rho^{\dagger} \sigma_{-}^{\left( j \right)} \sigma_{+}^{\left( j \right)} \right) \right. \notag \\
    &\quad + \left. \gamma_{j \alpha}^{-} \left( \sigma_{-}^{\left( j \right)} \rho^{\dagger} \sigma_{+}^{\left( j \right)} - \frac{1}{2} \sigma_{+}^{\left( j \right)} \sigma_{-}^{\left( j \right)} \rho^{\dagger} - \frac{1}{2} \rho^{\dagger} \sigma_{+}^{\left( j \right)} \sigma_{-}^{\left( j \right)} \right) \right] \notag \\
    &= \left[ -i \left( H_{s} \rho - \rho H_{s} \right) \right]^{\dagger} + \sum_{\alpha} \sum_{j = 1}^{N} \left[ \gamma_{j \alpha}^{+} \left( \sigma_{+}^{\left( j \right)} \rho \sigma_{-}^{\left( j \right)} - \frac{1}{2} \sigma_{-}^{\left( j \right)} \sigma_{+}^{\left( j \right)} \rho - \frac{1}{2} \rho \sigma_{-}^{\left( j \right)} \sigma_{+}^{\left( j \right)} \right) \right. \notag \\
    &\quad + \left. \gamma_{j \alpha}^{-} \left( \sigma_{-}^{\left( j \right)} \rho \sigma_{+}^{\left( j \right)} - \frac{1}{2} \sigma_{+}^{\left( j \right)} \sigma_{-}^{\left( j \right)} \rho - \frac{1}{2} \rho \sigma_{+}^{\left( j \right)} \sigma_{-}^{\left( j \right)} \right) \right]^{\dagger} \notag \\
    &= \left( \mathcal{L} \rho \right)^{\dagger} = \lambda^{*} \rho^{\dagger}.
\end{align}
Therefore, $\lambda^{*}$ is also an eigenvalue of $\mathcal{L}$ with the corresponding eigenmatrix $\rho^{\dagger}$.

\subsection{Eigenmatrices of a reduced Lindbladian}

Here, we provide the explicit expressions of the left- and right- eigenmatrices $\sigma_{i}$ and $\rho_{i}$ corresponding to the eigenvalues $\lambda_{i}$ of the reduce Lindbladian $\tilde{\mathcal{L}}$, see Eq.~\eqref{eq:reduced_L} and Ref.~\cite{khandelwal2021signatures}, away from exceptional points.
\begin{align}
    \rho_{ss} &= \frac{1}{\gamma^{2} \mathcal{G}_{0}} 
	\left( 
	\begin{array}{cc}
		4 g^{2} \gamma_{-}^{2} + \gamma_{1}^{-} \gamma_{2}^{-} \gamma^{2} & 0 \\
		0 & 4 g^{2} \gamma_{-} \gamma_{+} + \gamma_{1}^{-} \gamma_{2}^{+} \gamma^{2} \\
		0 & -i 2 g \left( \gamma_{1}^{-} \gamma_{2}^{+} - \gamma_{1}^{+} \gamma_{2}^{-} \right) \gamma \\
		0 & 0
	\end{array}
    \right. \notag \\
    &\hspace{18em}\left. 
    \begin{array} {cc}
        0 & 0 \\
        i 2 g \left( \gamma_{1}^{-} \gamma_{2}^{+} - \gamma_{1}^{+} \gamma_{2}^{-} \right) \gamma & 0 \\
        4 g ^{2} \gamma_{-} \gamma_{+} + \gamma_{1}^{+} \gamma_{2}^{-} \gamma^{2} & 0 \\
        0 & 4 g^{2} \gamma_{+}^{2} + \gamma_{1}^{+} \gamma_{2}^{+} \gamma^{2}
    \end{array}
	\right), \\
	\sigma_{1} &=
	\begin{pmatrix}
		1 & 0 & 0 & 0 \\
		0 & 1 & 0 & 0 \\
		0 & 0 & 1 & 0 \\
		0 & 0 & 0 & 1
	\end{pmatrix}, \\
    \rho_{2} &=
	\begin{pmatrix}
		1 & 0 & 0 & 0 \\ 
		0 & -1 & 0 & 0 \\ 
		0 & 0 & -1 & 0 \\ 
		0 & 0 & 0 & 1
	\end{pmatrix}, \\
	\sigma_{2} &= \frac{\gamma_{1}^{-} \gamma_{2}^{+} - \gamma_{1}^{+} \gamma_{2}^{-}}{\gamma^{2} \mathcal{G}_{0}}
	\left(
	\begin{array}{cc}
		\gamma_{1}^{+} \gamma - \gamma_{2} \gamma_{+} + \frac{\gamma_{+}^{2} \mathcal{G}_{0} }{\gamma_{1}^{-} \gamma_{2}^{+} - \gamma_{1}^{+} \gamma_{2}^{-}} & 0 \\
		0 & \gamma_{1}^{+} \gamma + \gamma_{2} \gamma_{-} - \frac{\gamma_{-} \gamma_{+} \mathcal{G}_{0}}{\gamma_{1}^{-} \gamma_{2}^{+} - \gamma_{1}^{+} \gamma_{2}^{-}} \\
		0 & -i 2 g \gamma \\
		0 & 0
	\end{array}
	\right. \notag \\
	&\hspace{10em} \left.
	\begin{array}{cc}
		0 & 0 \\
		i 2 g \gamma & 0 \\
		-\gamma_{1}^{-} \gamma - \gamma_{2} \gamma_{+} - \frac{\gamma_{-} \gamma_{+} \mathcal{G}_{0}}{\gamma_{1}^{-} \gamma_{2}^{+} - \gamma_{1}^{+} \gamma_{2}^{-}} & 0 \\
		0 & -\gamma_{1}^{-} \gamma + \gamma_{2} \gamma_{-} + \frac{\gamma_{-}^{2} \mathcal{G}_{0}}{\gamma_{1}^{-} \gamma_{2}^{+} - \gamma_{1}^{+} \gamma_{2}^{-}}
	\end{array}
	\right), \\
	\rho_{3} &=
	\begin{pmatrix}
		0 & 0 & 0 & 0 \\ 
		0 & 0 & 1 & 0 \\ 
		0 & 1 & 0 & 0 \\ 
		0 & 0 & 0 & 0 
	\end{pmatrix}, \\
	\sigma_{3} &=
	\begin{pmatrix}
		0 & 0 & 0 & 0 \\ 
		0 & 0 & \frac{1}{2} & 0 \\ 
		0 & \frac{1}{2} & 0 & 0 \\ 
		0 & 0 & 0 & 0
	\end{pmatrix}, \\
	\rho_{4} &=
	\begin{pmatrix}
		2 \gamma_{-} & 0 & 0 & 0 \\ 
		0 & \gamma_{+} - \gamma_{-} & -i \frac{\gamma \Delta\gamma}{2 g} & 0 \\ 
		0 & i \frac{\gamma \Delta\gamma}{2 g} & \gamma_{+} - \gamma_{-} & 0 \\ 
		0 & 0 & 0 & -2 \gamma_{+}
	\end{pmatrix}, \\
	\sigma_{4} &= \frac{g}{\gamma \eta^{2}}
	\begin{pmatrix}
		-\frac{4 g \gamma_{+}}{\gamma} & 0 & 0 & 0 \\ 
		0 & -\frac{2 g \left( \gamma_{+} - \gamma_{-} \right)}{\gamma} & -i \Delta\gamma & 0 \\ 
		0 & i \Delta\gamma & -\frac{2 g \left( \gamma_{+} - \gamma_{-} \right)}{\gamma} & 0 \\ 
		0 & 0 & 0 & \frac{4 g \gamma_{-}}{\gamma}
	\end{pmatrix}, \\
	\rho_{5} &=
	\left(
	\begin{array}{cc}
		\frac{\gamma_{-} \Delta\gamma + \left( \gamma_{1}^{-} - \gamma_{2}^{-} \right) \eta}{g} & 0 \\
		0 & \frac{\left( \gamma_{+} - \gamma_{-} \right) \Delta\gamma - 2 \eta^{2} - 2 \left( \gamma_{1}^{-} + \gamma_{2}^{+} \right) \eta}{2 g} \\
		0 & -i \left( \gamma + 2 \eta \right) \\
		0 & 0 
	\end{array}
    \right. \notag \\
    &\hspace{16em} \left. 
    \begin{array}{cc}
         0 & 0 \\
         i \left( \gamma + 2 \eta \right) & 0 \\
         \frac{\left( \gamma_{+} - \gamma_{-} \right) \Delta\gamma + 2 \eta^{2} + 2 \left( \gamma_{1}^{+} + \gamma_{2}^{-} \right) \eta}{2 g} & 0 \\
         0 & \frac{-\gamma_{+} \Delta\gamma - \left( \gamma_{1}^{+} - \gamma_{2}^{+} \right) \eta}{g}
    \end{array}
	\right), \\
	\sigma_{5} &= \frac{g}{4 \eta^{2} \mathcal{G}_{0}}
	\left(
	\begin{array}{cc}
		\gamma_{+} \Delta\gamma - \left( \gamma_{1}^{+} - \gamma_{2}^{+} \right) \eta & 0 \\
		0 & -\frac{1}{2} \left( \gamma_{-} - \gamma_{+} - 2 \eta \right) \left( \Delta\gamma + \eta \right) - 2 \gamma_{1}^{+} \eta \\
		0 & -i g \left( \gamma - 2 \eta \right) \\
		0 & 0
	\end{array}
	\right. \notag \\
	&\hspace{9em} \left.
	\begin{array}{cc}
		0 & 0 \\
		i g \left( \gamma - 2 \eta \right) & 0 \\
		-\frac{1}{2} \left( \gamma_{-} - \gamma_{+} - 2 \eta \right) \left( \Delta\gamma - \eta \right) + 2 \gamma_{2}^{+} \eta & 0 \\
		0 & -\gamma_{-} \Delta\gamma + \left( \gamma_{1}^{-} - \gamma_{2}^{-} \right) \eta
	\end{array}
	\right),\\
	\rho_{6} &=
	\left(
	\begin{array}{cc}
		\frac{\gamma_{-} \Delta\gamma - \left( \gamma_{1}^{-} - \gamma_{2}^{-} \right) \eta}{g} & 0 \\
		0 & \frac{\left( \gamma_{+} - \gamma_{-} \right) \Delta\gamma - 2 \eta^{2} + 2 \left( \gamma_{1}^{-} + \gamma_{2}^{+} \right) \eta}{2 g} \\
		0 & -i \left( \gamma - 2 \eta \right) \\
		0 & 0
	\end{array}
    \right. \notag \\
    &\hspace{16em} \left. 
    \begin{array}{cc}
         0 & 0 \\
         i \left( \gamma - 2 \eta \right) & 0 \\
         \frac{\left( \gamma_{+} - \gamma_{-} \right) \Delta\gamma + 2 \eta^{2} - 2 \left( \gamma_{1}^{+} + \gamma_{2}^{-} \right) \eta}{2 g} & 0 \\
         0 & \frac{-\gamma_{+} \Delta\gamma + \left( \gamma_{1}^{+} - \gamma_{2}^{+} \right) \eta}{g}
    \end{array}
	\right),\\
	\sigma_{6} &= \frac{g}{4 \eta^{2} \mathcal{G}_{0}}
	\left(
	\begin{array}{cc}
		\gamma_{+} \Delta\gamma + \left( \gamma_{1}^{+} - \gamma_{2}^{+} \right) \eta & 0 \\
		0 & -\frac{1}{2} \left( \gamma_{-} - \gamma_{+} + 2 \eta \right) \left( \Delta\gamma - \eta \right) + 2 \gamma_{1}^{+} \eta \\
		0 & -i g \left( \gamma + 2 \eta \right) \\
		0 & 0
	\end{array}
	\right. \notag \\
	&\hspace{9em} \left.
	\begin{array}{cc}
		0 & 0 \\
		i g \left( \gamma + 2 \eta \right) & 0 \\
		-\frac{1}{2} \left( \gamma_{-} - \gamma_{+} + 2 \eta \right) \left( \Delta\gamma + \eta \right) - 2 \gamma_{2}^{+} \eta & 0 \\
		0 & -\gamma_{-} \Delta\gamma - \left( \gamma_{1}^{-} - \gamma_{2}^{-} \right) \eta
	\end{array}
	\right).
\end{align}
One should be aware of that at exceptional point, $\rho_{4}$, $\rho_{5}$ and $\rho_{6}$ plus the corresponding right-eigenmatrices coalesce while $\rho_{3}$ and $\sigma_{3}$ do not.

\end{document}